# Quantification of predictive uncertainty in hydrological modelling by harnessing the wisdom of the crowd: A large-sample experiment at monthly timescale

Georgia Papacharalampous[1,*], Hristos Tyralis[2], Demetris Koutsoyiannis[3], and Alberto Montanari[4]

[1]  Department of Water Resources and Environmental Engineering, School of Civil Engineering, National Technical University of Athens, Heroon Polytechneiou 5, 157 80 Zographou, Greece; papacharalampous.georgia@gmail.com; https://orcid.org/0000-0001-5446-954X

[2]  Air Force Support Command, Hellenic Air Force, Elefsina Air Base, 192 00 Elefsina, Greece; montchrister@gmail.com; https://orcid.org/0000-0002-8932-4997

[3]  Department of Water Resources and Environmental Engineering, School of Civil Engineering, National Technical University of Athens, Heroon Polytechneiou 5, 157 80 Zographou, Greece; dk@itia.ntua.gr; https://orcid.org/0000-0002-6226-0241

[4]  Department of Civil, Chemical, Environmental and Materials Engineering (DICAM), University of Bologna, via del Risorgimento 2, 40136 Bologna, Italy; alberto.montanari@unibo.it; https://orcid.org/0000-0001-7428-0410

*  Correspondence: papacharalampous.georgia@gmail.com, tel: +30 69474 98589

**Abstract**: Predictive hydrological uncertainty can be quantified by using ensemble methods. If properly formulated, these methods can offer improved predictive performance by combining multiple predictions. In this work, we use 50-year-long monthly time series observed in 270 catchments in the United States to explore the performances provided by an ensemble learning post-processing methodology for issuing probabilistic hydrological predictions. This methodology allows the utilization of flexible quantile regression models for exploiting information about the hydrological model's error. Its key differences with respect to basic two-stage hydrological post-processing methodologies using the same type of regression models are that (a) instead of a single point hydrological prediction it generates a large number of "sister predictions" (yet using a single hydrological model), and that (b) it relies on the concept of combining probabilistic predictions via simple quantile averaging. A major hydrological modelling challenge is obtaining probabilistic predictions that are simultaneously reliable and associated to prediction bands that are as narrow as possible; therefore, we assess both these desired properties of the predictions by computing their coverage probabilities, average widths and average interval scores. The results confirm the usefulness of the proposed methodology and its larger robustness with respect to basic two-stage post-processing methodologies. Finally, this methodology is empirically proven to harness the "wisdom of the crowd" in terms of average interval score, i.e., the average of the individual predictions combined by this methodology scores no worse –usually better– than the average of the scores of the individual predictions.

**Key words**: ensemble learning; hydrological model; probabilistic prediction; quantile averaging; quantile regression; uncertainty quantification

## 1. Introduction

Uncertainty is a subject of ongoing discussions in hydrology (see e.g., Beven 1993, 2000, 2001; Vogel 1999; Beven and Feer 2001; Krzysztofowicz 2001a; Pappenberger and Beven 2006; Koutsoyiannis and Montanari 2007; Montanari 2007; Koutsoyiannis et al. 2009; Koutsoyiannis 2010, 2011; Kuczera et al. 2010; Ramos et al. 2010, 2013; Weijs et al. 2010; Juston et al. 2012; Nearing at al. 2016). Hydrological modelling uncertainty is traditionally recognised within the model calibration and validation phases (Montanari 2011) in the context of the widely accepted evaluation framework proposed by Klemeš (1986). Within this framework "uncertainty treatment" serves the verification of hydrological model's reliability (Montanari 2011). The large number of relevant studies and their high significance are summarised, for instance, in the review papers by Efstratiadis and Koutsoyiannis (2010), and Pechlivanidis et al. (2011).

As discussed in Koutsoyiannis (2010), an appropriate modelling approach for any uncertain hydrological system should necessarily include quantification of its uncertainty within a stochastic framework. Uncertainty is naturally quantified using the probability theory, i.e., in terms of probability distribution function (PDF; Todini 2007; see also Todini 2004, 2008). Todini (2007; quoting Krzysztofowicz 1999) emphasizes the fact that in engineering applications the targeted uncertainty quantification should be no other than the quantification of the predictive uncertainty, i.e., the total uncertainty of the predictand. Along with this strong engineering-oriented interest of hydrologists (which might be underestimated in some cases but is of vital significance for hydrology, as for any applied science; Shmueli 2010), understanding of predictive performance and uncertainty in hydrological modelling is undoubtedly a major science-oriented target (see e.g., Clark et al. 2008; Renard et al. 2010, 2011; Montanari 2011; Pechlivanidis et al. 2011; Beven 2012; Montanari and Koutsoyiannis 2012; Clark et al. 2015; Farmer and Vogel 2016; Széles et al. 2018; Khatami et al. 2019).

The preference for process-based (including conceptual) hydrological models (over the data-driven ones; Toth et al. 1999), along with both the practical relevance of predictive uncertainty quantification in hydrology and the attentiveness of hydrologists towards increasing understanding in (probabilistic) hydrological modelling, has led to the development of a wide range of methodologies for the integration of process-based and statistical models. This range includes (but is not limited to) various types of methodologies that statistically post-process the output of process-based models (hereafter referred to as "post-processing" methodologies). Considering information from deterministic models within uncertainty assessment frameworks (instead of exclusively using statistical methods) is a state-of-the-art methodological approach that is also adopted in contiguous fields (see e.g., Tyralis and Koutsoyiannis 2017). This approach holds a prominent position in the field of probabilistic hydrological modelling, in contrast to purely statistical probabilistic methodologies, which are rarely preferred; therefore, the below-provided outline exclusively focuses on it.

Perhaps the most frequently exploited methodology for predictive uncertainty quantification in hydrological modelling is the Generalized Likelihood Uncertainty Estimation (GLUE; Beven and Binley 2014). This approach has been proposed by Beven and Binley (1992), and is based on the concept of equifinality (see, e.g., Beven 2006; Khatami 2019). It has been discussed, for example, in Montanari (2005), Mantovan and Todini (2006), Stedinger et al. (2008), Vrugt et al. (2009b), and Sadegh and Vrugt (2013); see also the related comments in Todini (2007).

Another predictive uncertainty quantification methodology that has received attention both by researchers and practitioners is the Bayesian Forecasting System (BFS). The BFS has been introduced by Krzysztofowicz (1999, 2001a, 2002), Krzysztofowicz and Kelly (2000), and Krzysztofowicz and Herr (2001) for producing probabilistic river stage forecasts. It consists of three discrete components, namely the Precipitation Uncertainty Processor (PUB), the Hydrologic Uncertainty Processor (HUP) and the INTegrator (INT). Information about these components can be found in Kelly and Krzysztofowicz (2000), Krzysztofowicz and Kelly (2000), and Krzysztofowicz (2001b) respectively. This Bayesian methodology is conceived for real-time forecasting and relies on the assumption that uncertainty is mainly introduced by rainfall forecast errors.

There are also Bayesian post-processing methodologies that explicitly consider the contribution of input and output data uncertainty (which also affects the quantification of parameter uncertainty; see Coxon et al. (2015), Di Baldassarre et al. (2012), Di Baldassarre and Montanari (2009), Kauffeldt et al. (2013), McMillan et al. (2010), McMillan et al. (2012), Montanari and Di Baldassarre (2013), and Tomkins (2014) for information on rainfall-runoff data errors). Perhaps the most characteristic example of such a methodology is the Bayesian Total Error Analysis (BATEA) framework by Kavetski et al. (2002; see also Kavetski et al. 2006a, Kuczera et al. 2006), implemented, for instance, in Thyer et al. (2009) and Renard et al. (2010, 2011). This Bayesian framework facilitates the joint modelling of parameter uncertainty, data uncertainties, and model error, i.e., of all sources of uncertainty that are often assumed to collectively compose the predictive uncertainty. Other Bayesian post-processing methodologies introduced for parameter and predictive uncertainty quantification are described by Kuczera (1983), Schoups and Vrugt (2010), Evin et al. (2013; see also Evin et al. 2014), Hernández-López and Francés (2017) and Romero-Cuellar et al. (2019); see also the literature review in Hernández-López and Francés (2017).

Non-Bayesian post-processing methodologies that in their majority focus on the modelling of a single error term conditional on hydrological point predictions and historical information are also available in the hydrological modelling literature (see e.g., Bock et al. 2018; Bourgin et al. 2015; Farmer and Vogel 2016; Montanari and Brath 2004; Montanari and Grossi 2008; Dogulu et al. 2015; López López et al. 2014; Solomatine and Shrestha 2009; Wani et al. 2017). Adopting the terminology by Evin et al. (2014), such methodologies are hereafter referred to as "two-stage" post-processing methodologies, as their hydrological and error models are estimated in two subsequent stages. It is relevant to note at this point that Bayesian and two-stage post-processing methodologies are rather not directly comparable, since they are characterized by different statistical-modelling-culture traits and distinguishing features, which in their turn lead to different advantages and disadvantages (see Appendix A). For extensive discussions on the statistical modelling cultures, the reader is referred to Breiman (2001) and Shmueli (2010).

In the context described so far, Montanari and Koutsoyiannis (2012) introduced a flexible two-stage post-processing methodology (hereafter referred to as "MK blueprint methodology") that facilitates both probabilistic modelling and understanding from a stochastic perspective of rainfall-runoff (and other stochastic) relationships. In its basic configuration, this methodology utilizes a single hydrological model to generate a large number of point predictions (hereafter referred to as "sister predictions"; adopting a similar terminology to the one by Nowotarski et al. 2016, Wang et al. 2016, and Liu et al. 2017). As implied by its post-processing nature, it also utilizes a second –necessarily statistical– model for modelling the error of the hydrological model (hereafter referred to as "error model").

Different variants of the MK blueprint methodology can be found in Sikorska et al. (2015), Quilty et al. (2019) and Papacharalampous et al. (2019b; companion to the present paper). The original blueprint and the variant by Sikorska et al. (2015) are formulated to explicitly consider input data uncertainty, while in both related papers a large number of hydrological model parameters are obtained by using the DREAM algorithm by Vrugt et al. (2009a; see also Vrugt 2016). This algorithm (see, e.g., Schoups and Vrugt 2010; Laloy and Vrugt 2012; Vrugt et al. 2013; Sadegh and Vrugt 2014) is a popular Markov chain Monte Carlo (MCMC) algorithm for sampling from the posterior parameter distribution of hydrological models (see also the related implementations in Sadegh et al. 2015; Hernández-López and Francés 2017; Vrugt et al. 2008; Volpi et al. 2017). Other (non-Bayesian) methodologies could also be used for obtaining a large number of hydrological model parameters (Montanari and Koutsoyiannis 2012), while in absence of relevant information the MK blueprint methodology can also be applied without explicitly considering input data uncertainty (see e.g., the implementations in Quilty et al. 2019 and the formulations of the variants in Papacharalampous et al. 2019b). Quilty et al. (2019) perform probabilistic water demand forecasting using exogenous variables; therefore, their variants constitute integrations within the MK blueprint framework of concepts particularly useful and/or popular for this task, such as bootstrapping, variable selection and wavelet decomposition.

In spite of their (larger or smaller) differences in terms of conceptualization, underlying modelling cultures and inherent modelling assumptions, all the above-outlined state-of-the-art techniques aim at filling a common knowledge gap that currently exists in the probabilistic hydrological modelling and forecasting literatures, specifically at answering the following research question: How to reduce modelling uncertainty as much as possible? Risk reduction in (probabilistic) hydrological modelling is the 20th of the 23 major "unsolved" hydrological problems, as posed by Blöschl et al. (2019, Section 3) through a community-based process. The present study aspires to contribute to the large efforts made towards solving this problem.

We extensively test the hydrological modelling capabilities provided by the variants of the MK blueprint methodology introduced in Papacharalampous et al. (2019b) (hereafter collectively referred to as "working methodology"), when these variants are applied using the quantile regression model by Koenker and Bassett (1978; see also Koenker 2005) as error model. The quantile regression model is a balanced choice between interpretable and more flexible algorithms from the statistical learning literature. It has already been applied for post-processing hydrological predictions within hydrological modelling case studies (see e.g., Dogulu et al. 2015, López López et al. 2014, Solomatine and Shrestha 2009, Wani et al. 2017), while its use is more common in the field of hydrological forecasting (see e.g., Tyralis et al. 2019a and the references therein); see also the references in Dogulu et al. (2015), and Abbas and Xuan (2019) for applications of this model in other geoscience concepts.

For benchmarking purposes, we also apply the working methodology using the linear regression model (see e.g., James et al. 2013; Hastie et al. 2009) as error model, and two naïve probabilistic data-driven schemes. For the merits of using benchmarks in hydrological modelling, the reader is referred to Pappenberger et al. (2015); see also benchmarking examples in Montanari and Brath (2004), Papacharalampous and Tyralis (2018), Papacharalampous et al. (2018a,b,c, 2019a,b,d), Quilty et al. (2019), Evin et al. (2014), Sikorska et al. (2015), Tyralis and Papacharalampous (2017, 2018), Tyralis et al. (2018, 2019a,c), and Xu et al. (2018).

The working methodology is implemented within a large-sample real-world experiment. In the latter, we probabilistically solve monthly rainfall-runoff modelling problems for 270 catchments in the United States (US). Large-sample hydrological studies are increasingly carried out in the literature (see e.g., Bock et al. 2018; Bourgin et al. 2015; Coxon et al. 2015; Farmer and Vogel 2016; Langousis et al. 2016; Mouelhi et al. 2006a,b; Papacharalampous et al. 2018a,b, 2019a,d; Papalexiou and Koutsoyiannis 2013; Perrin et al. 2001; Ren et al. 2016; Sawicz et al. 2011; Tyralis and Koutsoyiannis 2017; Tyralis and Papacharalampous 2017, 2018; Tyralis et al. 2018, 2019a,c; Weijs et al. 2013; Xu et al. 2018, 2019), while this is the first study performing a large-scale assessment of the MK blueprint methodology.

The aims of the study (that can be addressed only within a large-sample hydrological study) are to:

1) Validate the working methodology.

2) Compare its variants both in terms of predictive performance and computational requirements.

3) Quantify the improvement in performance when using the quantile regression model instead of the linear regression model as error model. In contrast to the latter model, the former model is known to be appropriate for modelling heteroscedasticity (Koenker and Hallock 2001; Koenker 2005).

4) Demonstrate in real-world applications the larger robustness in performance of the working methodology compared to two-stage post-processing methodologies producing a single point hydrological prediction (hereafter referred to as "basic" two-stage post-processing methodologies).

5) Provide an empirical proof of the ability of the working methodology to harness the wisdom of the crowd. This ability stems from the concept of combining probabilistic predictions via simple quantile averaging, on which this methodology relies, while in Lichtendahl et al. (2013, Section 5) it is defined as follows: The average of predictions scores no worse –usually better– than the average of the scores of the combined predictions. According to the same study, this ability has to be empirically proven for the problem and scores of interest, since the proofs in Lichtendahl et al. (2013) are made for stylized versions.

## 2. Data and methods

In this section, we present the experimental methodology of the study by emphasizing implementation details, as it is suggested by the guidelines by Abrahart et al. (2008). Statistical software information is summarized in Appendix B. The working methodology is outlined in Appendix C, while the reader is referred to Papacharalampous et al. (2019b) for its detailed and formal presentation.

### 2.1 Rainfall-runoff dataset

We use the US Model Parameter Estimation Experiment (MOPEX) dataset, which is documented in Schaake et al. (2006; see also Schaake et al. 2000, Duan et al. 2006, Wagener et al. 2006). This dataset comprises hydrometeorological and land-surface-characteristic data originating from US catchments of intermediate size, and has been extensively used in hydrological studies (see e.g., Kavetski et al. 2006b; Sawicz et al. 2011; Huang et al. 2013; Evin et al. 2014; Weijs et al. 2013; Ye et al. 2014; Ren et al. 2016; Hernández-López and Francés 2017). All included catchments are unregulated; therefore, the modelling assumption of stationarity is reasonable on these real-world data (see e.g., Koutsoyiannis 2011; Montanari and Koutsoyiannis 2014; Koutsoyiannis and Montanari 2015).

From the original dataset we retrieve daily information about mean areal precipitation, climatic potential evaporation and streamflow discharge for 431 US catchments. The retrieved data span from January 1st, 1948 to December 31st, 2003, thus covering a 56-year period, yet containing a large amount of missing and negative (unrealistic) values. We process the retrieved data aiming to simultaneously achieve two objectives, i.e., (a) extracting time series blocks covering a long common period of complete historical information (with no missing or unreliable values), and (b) retaining historical information for a large number of catchments. A satisfactory compromise between these two objectives is reached when using as sampling period each of the periods 1950–1999 and 1949–1998. Both these samplings result in 50 (calendar) years of complete daily time series data for 270 catchments. We adopt the former option, as it offers (slightly) more recent data compared to the alternative one. The retained time series data are aggregated to produce total monthly precipitation, potential evaporation and streamflow discharge time series, each comprising 600 values. The resulted total monthly time series constitute the herein examined dataset. The locations of the examined MOPEX catchments are depicted in Figure 1. A wide range of climate regimes is well-represented by this sample set of catchments (see Kottek et al. 2006).

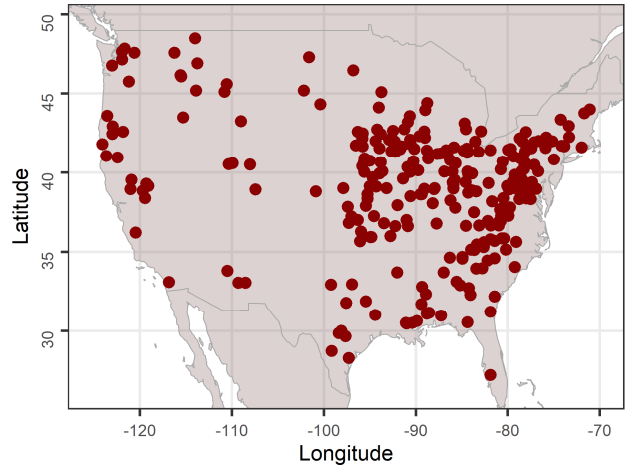

Figure 1. Locations of the 270 MOPEX catchments examined within the large-sample experiment of the study. The data are sourced from Schaake et al. (2006).

## 2.2 Prediction interval obtainment

### 2.2.1 Overview of modelling methodology

The monthly data of Section 2.1 are handled as described in Section 2.2.2. We use these data to assess two basic and six ensemble schemes in obtaining interval predictions. Two statistical learning regression models (see Section 2.2.3) and one hydrological model (see Section 2.2.4) are utilized for this assessment. We define the prediction problem to be solved as the problem of predicting the quantiles with probability $p \in \{0.005, 0.0125, 0.025, 0.05, 0.10, 0.90, 0.95, 0.975, 0.9875, 0.995\}$ of monthly streamflow discharge in the period $T_3$ (hereafter referred to as "quantiles of interest") given monthly precipitation and monthly potential evaporation observations for the period $\{T_0, T_1, T_2, T_3\}$ and monthly streamflow discharge observations for the period $\{T_0, T_1, T_2\}$. These periods are defined in Section 2.2.2.

The basic schemes are "linear regression" and "quantile regression". Both of them are implemented by training the regression model directly on monthly data for the period $\{T_0, T_1, T_2\}$ and, subsequently, by using the trained regression model to predict the quantiles of interest (for the period $T_3$). The predictor variables in regression are monthly precipitation at time $t$ and monthly potential evaporation at time $t$, while the response variable is monthly streamflow discharge at time $t$. We note that these benchmark implementations of the regression models can only be viewed as naïve data-driven approaches to probabilistic hydrological modelling (because of the small number of predictor variables utilized). For more sophisticated implementations (which are outside of the scope of the study), more predictor variables could be used.

On the other hand, the ensemble schemes can be perceived as different configurations of the working methodology (allowing us to address the aims of the study). Ensemble schemes 1–3 (4–6) are based on variants 1–3 respectively of this methodology. Moreover, ensemble schemes 1–3 utilize a different statistical learning regression model as error model with respect to ensemble schemes 4–6. Specifically, ensemble schemes 1–3 utilize the linear regression model, while ensemble schemes 4–6 utilize the quantile regression model. The same ensemble schemes are also implemented in Papacharalampous et al. (2019b); however, their implementation therein is made by using toy hydrological models.

We describe here below the application of the ensemble schemes for a single catchment; the extension to all catchments is straightforward. The following steps are made once for all ensemble schemes:

1) We use monthly precipitation, potential evaporation and streamflow discharge observations for the period $T_1$ to obtain 600 sets of the hydrological model's parameters, as detailed in Section 2.2.4. This number of parameter sets offers a good compromise between computational requirements and predictive performance. We use these parameters to define 600 sister model realizations.

2) We obtain 600 sister predictions for the period $\{T_2, T_3\}$. Each sister prediction is obtained by implementing a different sister model realization given the monthly precipitation and potential evaporation observations for the same period. Each sister prediction contains 444 values.

3) We compute the sister model realizations' errors in the period $T_2$ by using the parts of the sister predictions extending in the same period alongside with their corresponding target values. The total number of the computed error values is 600 × 144 = 86 400.

The following steps are made independently by each ensemble scheme:

4) We train the error model in the period $T_2$. Specifically, we regress the sister model realizations' error at time $t$ (response variable) on the sister prediction at time $t$ (predictor variable). Ensemble schemes 1 and 4 train the error model 600 times, each time using a different sister prediction and its corresponding sister model realization's errors (use of 600 training datasets of size 144). Ensemble schemes 2 and 5 train the error model once by using all sister predictions and their corresponding sister model realizations' errors (use of one training dataset of size 86 400). Ensemble schemes 3 and 6 train the error model once by using a randomly selected sister prediction and its corresponding sister model realization's errors (use of one training dataset of size 144). The result of this step is 600 trained versions of the error model (each corresponding to a specific sister prediction) for each of the ensemble schemes 1 and 4, and one trained version of the error model for each of the ensemble schemes 2, 3, 5 and 6.

5) We apply the trained versions of the error models, obtained in the preceding step, to predict the quantiles with probability $p \in \{0.005, 0.0125, 0.025, 0.05, 0.10, 0.90, 0.95, 0.975, 0.9875, 0.995\}$ of each sister model realization's errors in the period $T_3$ given their corresponding sister prediction. For each ensemble scheme, the result of this step is 600 probabilistic predictions, each consisting of 10 quantile predictions.

6) We obtain 600 auxiliary probabilistic predictions of the process of interest, each consisting of 10 quantile predictions, by subtracting each of the 600 × 10 = 6 000 quantile predictions from its corresponding sister prediction.

7) The finally delivered predictive quantile with probability $p \in \{0.005, 0.0125, 0.025, 0.05, 0.10, 0.90, 0.95, 0.975, 0.9875, 0.995\}$ at time $t \in T_3$ is the average over all auxiliary predictive quantiles with the same probability $p$ at time $t$, i.e., the average of 600 in number auxiliary predictive quantiles. The finally delivered predictive quantiles of the process of interest form the 99%, 97.5%, 95%, 90% and 80% central prediction intervals.

The total number of sister predictions produced herein is 270 × 600 = 162 000, each containing 444 values, while the total number of auxiliary quantile predictions is 270 × 600 × 10 × 6 = 9 720 000, each containing 300 values, and the finally delivered quantile predictions are 270 × 10 × 8 = 21 600, each containing 300 values. For addressing aim 2 of the study, we measure the computational time consumed by each ensemble scheme.

### 2.2.2 Data handling and related remarks

Following the notations provided in Appendix C, we define the periods $T_1 = \{13, ..., 156\}$, $T_2 = \{157, ..., 300\}$ and $T_3 = \{301, ..., 600\}$ (corresponding to years 1951–1962, 1963–1974 and 1975–1999 respectively). We include a large amount of the available information in the period $T_3$ to facilitate proper testing. We also define period $T_0$ = {1, …, 12} (corresponding to year 1950). This period is used for warming-up the hydrological model (see Section 2.2.4). One-year warming-up periods are often assumed adequate for achieving an optimal state initialisation, while also allowing the full exploitation of the available historical information (see e.g., Edijatno et al. 1999; Perrin et al. 2003; Kim et al. 2018; see also the implementations in Xu 2001; Perrin et al. 2001; Mouelhi et al. 2006b; Vrugt et al. 2008).

We note that the data are used without any transformation applied to it. We attempted to apply the linear regression and quantile regression schemes to river discharge data that were pre-processed by using the square-root transformation. Nevertheless, this pre-processing (not presented here for reasons of brevity) had a negative effect on the quality of the naïve probabilistic predictions, mainly to those delivered by the linear regression scheme; therefore, it was abandoned. Moreover, a logarithmic transformation was not feasible, due to some zero monthly values of river discharge. We also attempted to apply the Yeo-Johnson and ordered quantile normalization transformations on the response, when solving the error modelling problems outlined in Section 2.2.1 (steps 4–5 of the application of the ensemble schemes). These transformations were also abandoned due to infinite predicted values. The square-root and logarithmic transformations on the response variable, i.e., the error of the hydrological model at time $t$, are not feasible due to the existence of negative error values.

### 2.2.3 Regression models and related procedures

We implement the linear regression and quantile regression models. Koenker and Hallock (2001) comprehensively discuss the difference in rationale behind these two models, as summarized subsequently. The training outcome in linear regression (i.e., least-squares regression with i.i.d. Gaussian errors with zero mean and constant variance; James et al. 2013) is a conditional mean function. The latter is a function describing how the mean of the response variable changes with the changes of the predictor variables. This function is obtained by minimizing a sum of squared residuals. On the contrary, the training outcome in quantile regression is a set of conditional quantile functions, obtained by minimizing the average quantile score. While in linear regression the PDF of the response variable is assumed to have the exact same variance and distributional shape independently of the values of the predictors, quantile regression does not make any particular assumption about this PDF; therefore, allowing a more representative description of the relationship between predictors and predictand. We use these two models to solve the regression problems described in Section 2.2.1. We train the quantile regression model by implementing the training algorithm by Koenker and d'Orey (1987, 1994).

### 2.2.4 Hydrological model and related procedures

We implement the monthly GR2M model by Mouelhi et al. (2006b), a parsimonious lumped conceptual model comprising only two parameters, that has been widely applied in the literature (see e.g., Paturel et al. 1995; Niel et al. 2003; Huard and Mailhot 2008; Louvet et al. 2016). This model was developed by adopting a stepwise procedure aiming to identify the most useful components of a five-parameter model. The latter was inspired from the structures of the monthly model by Makhlouf and Michel (1994), and the daily GR4J model by Perrin et al. (2003; see also Edijatno et al. 1999, Perrin et al. 2001). The first parameter ($\theta_1$) is the maximum capacity of the soil moisture reservoir expressed in mm, while the second one ($\theta_2$) represents water exchange between the studied and adjacent catchments. Values of the second parameter larger (smaller) than 1 indicate water supply from (to) adjacent catchment(s).

We simulate the posterior distribution of the parameters of the GR2M model conditional on the observations of the period $T_1$ within a Bayesian MCMC framework. We use flat priors for both the parameters $\theta_1$ and $\theta_2$. The likelihood error function is defined by Equation (1), where $y_t$ is the monthly streamflow discharge observations at time $t$, $u_t(\theta_1, \theta_2)$ is the prediction of the GR2M model at time $t$ and $|T_1|$ is the number of target data points included in the period $T_1$. We run 3 parallel Markov chains with different initial values, each comprising 2 000 iterations. The iterative simulation is performed by using the DRAM algorithm by Haario et al. (2006).

$$L(\theta_1, \theta_2) \propto \left(\sum_t (y_t - u_t(\theta_1, \theta_2))^2\right)^{-|T_1|/2} \quad (1)$$

We assess the approximate convergence of these chains by implementing the algorithm of Brooks and Gelman (1998), i.e., a multivariate version of the algorithm of Gelman and Rubin (1992). Amongst the outputs of this algorithm is a point estimate that is assumed to be informative about the approximate convergence, while it is based on a comparison of within-chain and between-chain variances. Point estimates substantially larger than 1 indicate lack of convergence. The simulation process is repeated until a point estimate smaller than 1.10 is delivered. Once the simulation is over, we retain the last 200 values of each chain, i.e., 600 values in total for each catchment. An example of simulated and retained parameters is presented in Figure 2.

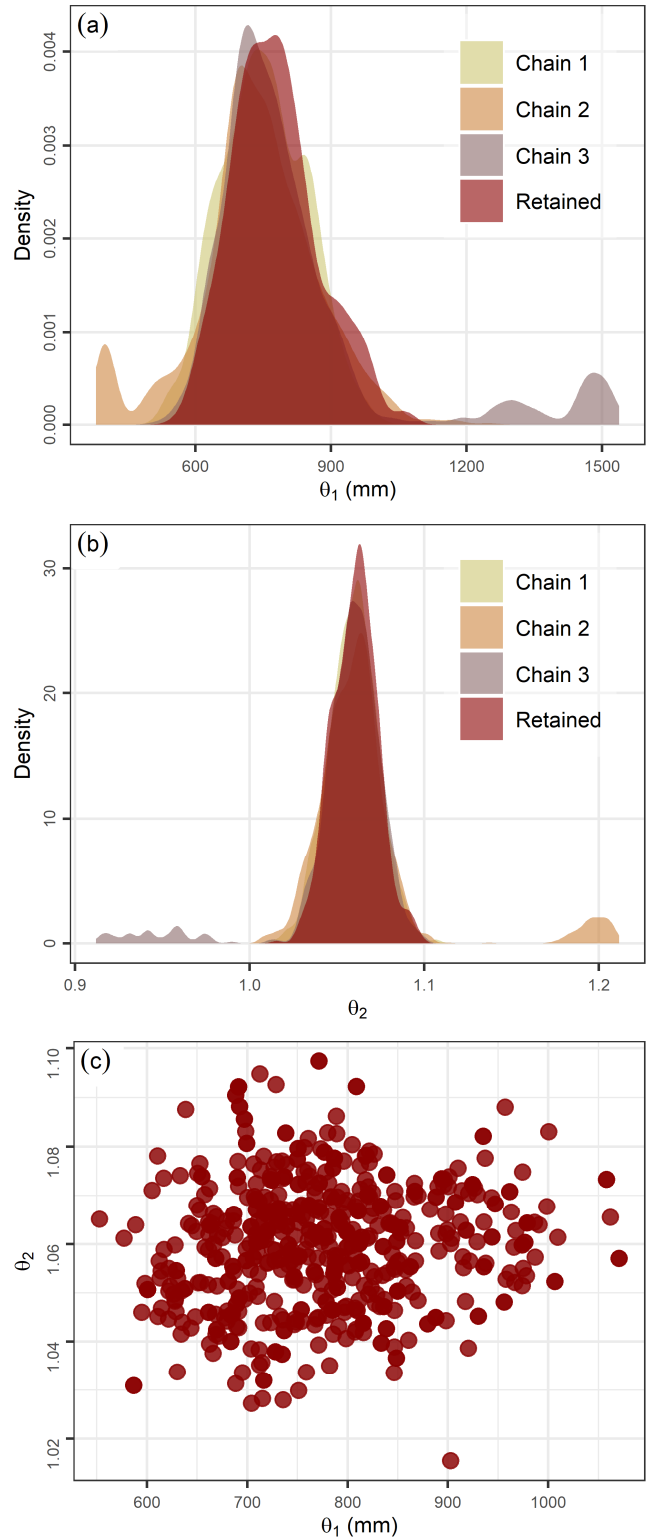

Figure 2. Simulated chains in (a–b), and retained parameter values in (a–c) obtained using precipitation, potential evaporation and streamflow discharge information for the period $T_1$ (years 1951–1962) for a randomly selected catchment.

## 2.3 Prediction interval assessment

We assess the quality of the interval predictions by computing their coverage probabilities, average widths and average interval scores. These metrics are used according to Table 1 to assess two desired properties in probabilistic modelling, i.e., the reliability and sharpness of interval predictions. The former property is defined as the statistical correspondence between the probabilistic forecasts and the observations, while the latter is the concentration of the predictive PDFs in absolute terms (Gneiting and Katzfuss 2014; see also Gneiting et al. 2007; Gneiting and Raftery 2007). For illustrative purposes, we also present examples of prediction intervals. We do not present QQ-plots for the following two reasons: i) we deliver predictive quantiles with probabilities that are either equal or smaller than 0.10, or equal or larger than 0.90 (since we are interested in specific prediction intervals; see Section 2.2.1), while QQ-plots are ideal when PDF predictions (or at least sets of predictive quantiles with probabilities running on a grid from 0 to 1) are delivered, and ii) we are interested in objectively assessing on a massive scale the predictive performance of several prediction schemes (separately for each of them) in 270 catchments, while QQ-plots are particularly useful for assessments made on a smaller scale.

Table 1. Metrics used for assessing the prediction interval $(1 - \alpha)$, $0 < \alpha < 1$.

| Metric | Definition | Possible values | Preferred values | Criterion/criteria |
|---|---|---|---|---|
| Coverage probability ($CP_\alpha$) | Equation (2) | [0, 1] | Smaller $|CP_\alpha - (1 - \alpha)|$ | Reliability |
| Average width ($AW_\alpha$) | Equation (3) | [0, +∞) | Smaller $AW_\alpha$ | Sharpness |
| Average interval score ($AIS_\alpha$) | Equation (4) | [0, +∞) | Smaller $AIS_\alpha$ | Reliability, sharpness |

For a specific central prediction interval of level $(1 - \alpha)$, $0 < \alpha < 1$, extending in the period $T_3$, the coverage probabilities, average widths and average interval scores are defined with Equations (2–4) respectively, where $v_{p,t}$ is the predictive quantile with probability $p \in \{\alpha/2, 1 - \alpha/2\}$ of monthly streamflow discharge at time $t$, I(·) is the indicator function and $|T_3|$ is the number of the target data points included in the period $T_3$.

$$CP_\alpha := \sum_t I(y_t \in [v_{(\alpha/2),t}, v_{(1 - \alpha/2),t}])/|T_3| \quad (2)$$

$$AW_\alpha := \sum_t (v_{(1 - \alpha/2),t} - v_{(\alpha/2),t})/|T_3| \quad (3)$$

$$AIS_\alpha := \sum_t ((v_{(1 - \alpha/2),t} - v_{(\alpha/2),t}) + (2/\alpha)(v_{(\alpha/2),t} - y_t) I(y_t < v_{(\alpha/2),t}) + (2/\alpha)(y_t - v_{(1 - \alpha/2),t}) I(y_t > v_{(1 - \alpha/2),t}))/|T_3| \quad (4)$$

Some remarks should be made on the (average) interval score. This score is appropriate for assessing probabilistic predictions in the form of prediction intervals (Gneiting and Raftery 2007, Section 6.2). It has three components (see Equation 4 above). The first component is the width of the prediction interval. As smaller values of the (average) interval score indicate better predictions than larger values (for a specific prediction problem), this component penalizes more the wider prediction intervals than the narrower ones (thereby rewarding narrow prediction intervals). The two remaining components quantify the distance between each of the two predictive quantiles forming the prediction interval and the observed value, in case that the latter falls outside of the prediction interval, and penalize larger distances more than smaller distances. In general, the (average) interval score should become smaller as we move from the outer to the inner prediction intervals. The reader is referred to Gneiting and Raftery (2007, Section 6.2) for detailed information on how to interpret this score.

Since the magnitude of the average interval score largely depends on the examined dataset, we mostly base our conclusions on relative improvements in terms of average interval score. The relative improvement in terms of average interval score, obtained when using a prediction interval $P_1$ (provided by a predictor of interest) with respect to another prediction interval $P_2$ of the same level (provided by a benchmark predictor), and denoted with $RI_{P_1,P_2}$, is computed according to Equation (5). In this equation, $AIS_{P_1}$ and $AIS_{P_2}$ denote the average interval scores of prediction interval $P_1$ and prediction interval $P_2$ respectively when they are computed over the whole time series; see Equation (4).

$$RI_{P_1,P_2} := (AIS_{P_2} - AIS_{P_1})/AIS_{P_2} \qquad (5)$$

Specifically, for addressing aims 1–3 of the study we compute the relative improvements provided all prediction schemes with respect to the linear regression and quantile regression schemes, and the relative improvements provided by ensemble schemes 4–6 with respect to ensemble schemes 1–3. For addressing aim 4 of the study, we use all auxiliary quantile predictions (9 720 000 in number) and the finally delivered quantile predictions (21 600 in number) to compute the relative improvements in terms of average interval score, when using the output of each ensemble scheme instead of each of the auxiliary interval predictions combined to obtain this output, according to Equation (6). In this equation, $AIS_{OUT}$ denotes the average interval score of the output interval prediction (obtained by using the method), $AIS_{IN_i}$ the average interval score of one from the auxiliary interval predictions {$IN_i$, $i = 1, ..., 600$} that are averaged by the method to obtain the output interval prediction (with average interval score equal to $AIS_{OUT}$), and $RI_{OUT,IN_i}$ the relative improvement of interest.

$$RI_{OUT,IN_i} := (AIS_{IN_i} - AIS_{OUT})/AIS_{IN_i} \qquad (6)$$

Finally, for addressing aim 5 of the study we use the same quantile predictions used for addressing aim 4 to compute the relative differences between the average interval score computed for the outputs of the ensemble schemes, i.e., the average of 600 probabilistic predictions (denoted with $AIS_{OUT}$; see above), and the average of the average interval scores computed for each of the combined auxiliary interval predictions $\{AIS_{IN_i}, i = 1, ..., 600\}$ (denoted with $AAIS_{IN}$; see also Equation (7) for its definition), the latter used as reference for the former. The computation of these relative differences is made using an equation analogous to Equations (5) and (6) above, specifically Equation (8), where $RD_{OUT,AAIS_{IN}}$ denotes the relative difference of interest.

$$AAIS_{IN} := \sum_{i=1}^{600}(AIS_{IN_i})/600 \quad (7)$$

$$RD_{OUT,AAIS_{IN}} := (AAIS_{IN} - AIS_{OUT})/AAIS_{IN} \quad (8)$$

## 3. Results and discussions

### 3.1 Addressing aims 1–3 of the study

This section is devoted to addressing aims 1–3 of the study. The presentation is mostly made in an aggregated form across all the examined catchments, while emphasis is placed on the average interval scores computed for the obtained prediction intervals and on the relative improvements provided by the ensemble schemes with respect to the basic schemes in terms of the same metric. This choice is implied by the fact that an objective co-assessment regarding reliability and sharpness provided, for instance, by the interval score is of the most practical relevance in technical applications; for a justification see Papacharalampous et al. (2019b); see also Gneiting and Katzfuss (2014). In spite of this placed emphasis and keeping pace with studies, such as those of Renard et al. (2010, 2011), Evin et al. (2013, 2014), Papacharalampous et al. (2019d) and Tyralis et al. (2019a), we start the presentation by separately summarizing the information that is purely related to the assessment of reliability from the information that is purely related to the assessment of sharpness. In this way, we facilitate an adequate degree of interpretability and understanding of what follows.

In Figure 3, we present several examples of prediction intervals, all delivered by ensemble scheme 5, in comparison to the targeted data points. As extracted from Figure 3, this scheme offers a (rather) high degree of reliability, i.e., it delivers prediction intervals that mostly contain the desired percentage of data points. The same applies to the remaining prediction schemes. Herein the related information is objectively summarized with Figure 4 and Table 2. In Figure 4, we comparatively present the boxplots of the coverage probabilities computed for all delivered and assessed solutions to the 270 examined rainfall-runoff problems. These coverage probabilities are rather good (than bad). The latter characterization holds, especially if we consider that the examined monthly time series are of only 600 values. In particular, the coverage probabilities for the 95% prediction intervals are comparable to those computed for the probabilistic predictions of Bock et al. (2018).

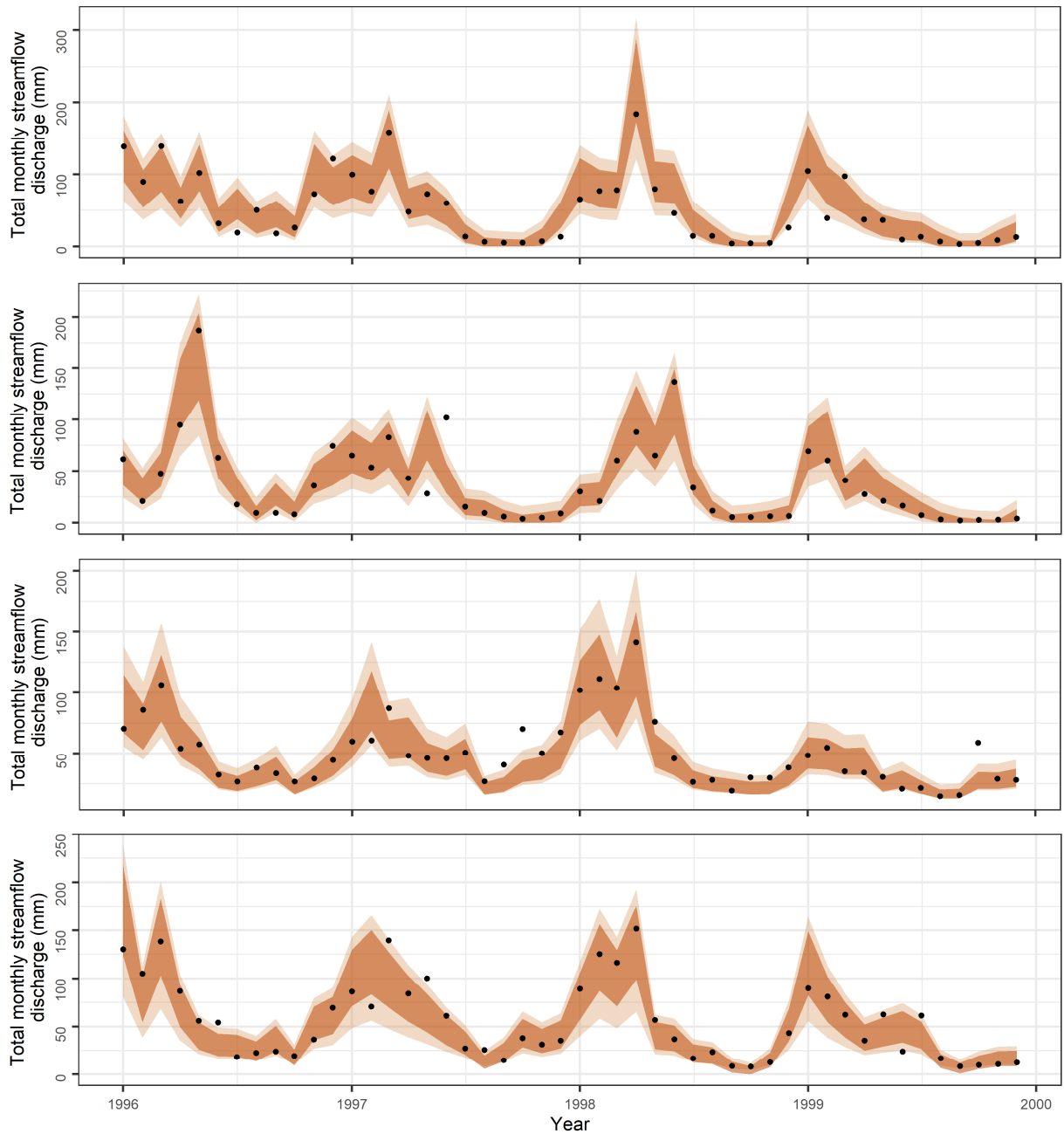

Figure 3. Prediction intervals provided by ensemble scheme 5 for four arbitrary catchments and a common 4-year sub-period of the period $T_3$ (years 1996–1999). Black dots denote the targeted points, while light orange and dark orange ribbons denote the 95% and 80% prediction intervals respectively.

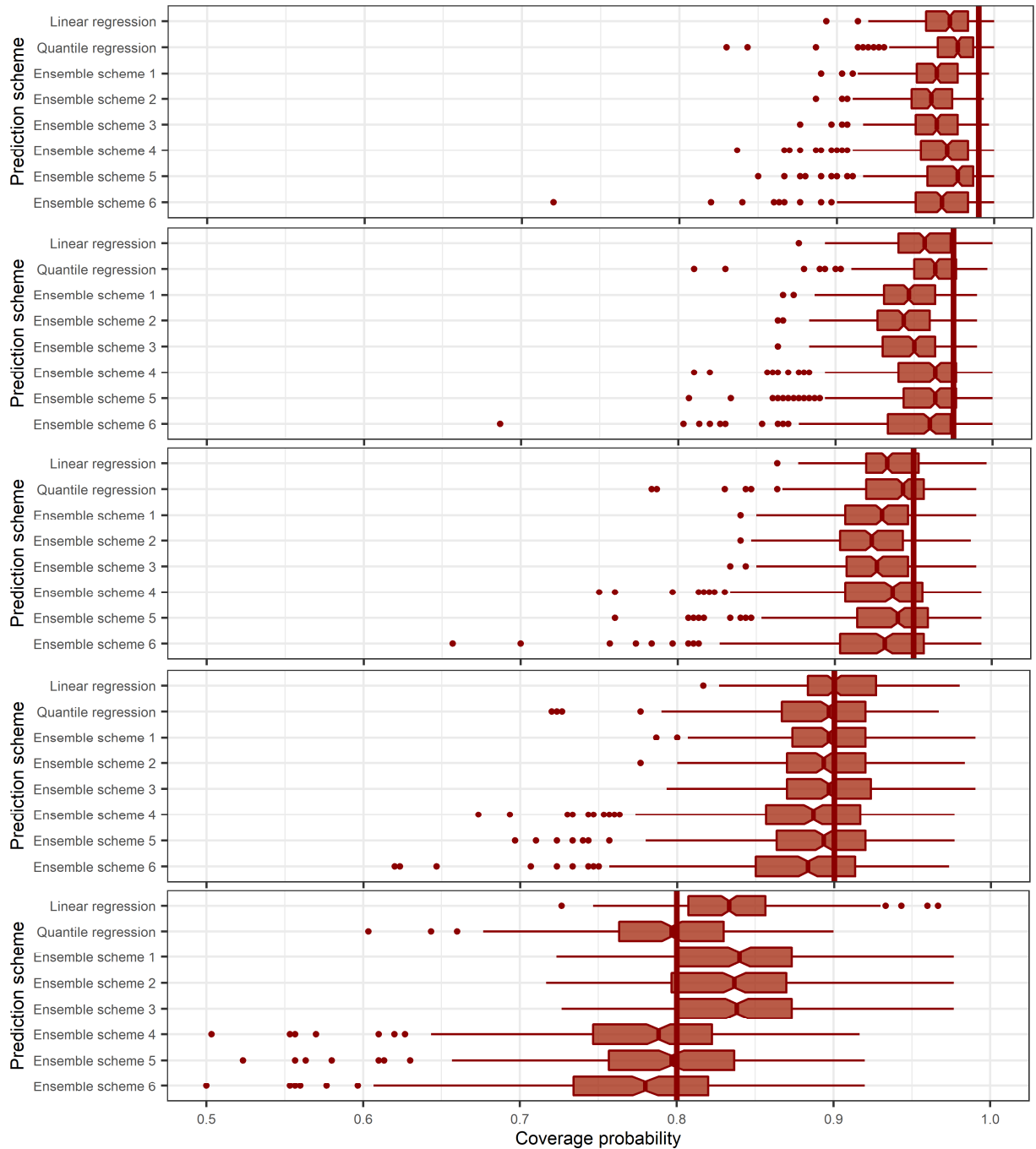

Figure 4. Coverage probabilities computed for the 99%, 97.5%, 95%, 90% and 80% prediction intervals (from top to bottom) delivered by the compared schemes for the period $T_3$ (years 1975–1999). Each boxplot summarizes 270 values. The optimal values are denoted with red thick vertical lines.

Table 2. Average coverage probabilities computed for the prediction intervals delivered by the compared schemes for the period $T_3$ (years 1975–1999). Each presented value summarizes 270 metric values.

| Prediction scheme | 99% prediction intervals | 97.5% prediction intervals | 95% prediction intervals | 90% prediction intervals | 80% prediction intervals |
|---|---|---|---|---|---|
| Linear regression | 0.969 | 0.955 | 0.937 | 0.904 | 0.835 |
| Quantile regression | 0.973 | 0.961 | 0.936 | 0.889 | 0.793 |
| Ensemble scheme 1 | 0.962 | 0.946 | 0.926 | 0.895 | 0.834 |
| Ensemble scheme 2 | 0.959 | 0.943 | 0.923 | 0.892 | 0.834 |
| Ensemble scheme 3 | 0.962 | 0.946 | 0.926 | 0.895 | 0.837 |
| Ensemble scheme 4 | 0.965 | 0.953 | 0.928 | 0.881 | 0.781 |
| Ensemble scheme 5 | 0.969 | 0.956 | 0.932 | 0.886 | 0.789 |
| Ensemble scheme 6 | 0.961 | 0.948 | 0.923 | 0.874 | 0.773 |

While the average-case reliability of all prediction schemes is remarkably high (see Table 2), the performance of the prediction schemes in terms of coverage probabilities varies from catchment to catchment (see Figure 4). The observed differences in performance become larger, e.g., in terms of interquartile range of the formed datasets, as we move from the 99% to the 80% prediction intervals. Moreover, although differentiations are observed between prediction schemes, the overall performance of most schemes is rather of the same quality (in particular for the outer prediction intervals), with the quantile regression scheme and ensemble scheme 5 to be the best-performing, especially the former one.

The average widths, on the other hand, clearly favour the ensemble schemes over the basic schemes (see Figure 5), with ensemble schemes 4–6 providing sharper predictions than ensemble schemes 1–3. In terms of the same criterion, ensemble schemes from the former (latter) category exhibit remarkably close performance to each other. The same applies in terms of coverage probabilities. As already expected because of the large differences observed in the river discharge regimes of the examined catchments, the average widths of the prediction intervals may differ significantly from catchment to catchment. These differences become smaller, as we move from the outer to the inner prediction intervals, i.e., from the 99% to the 80% prediction intervals.

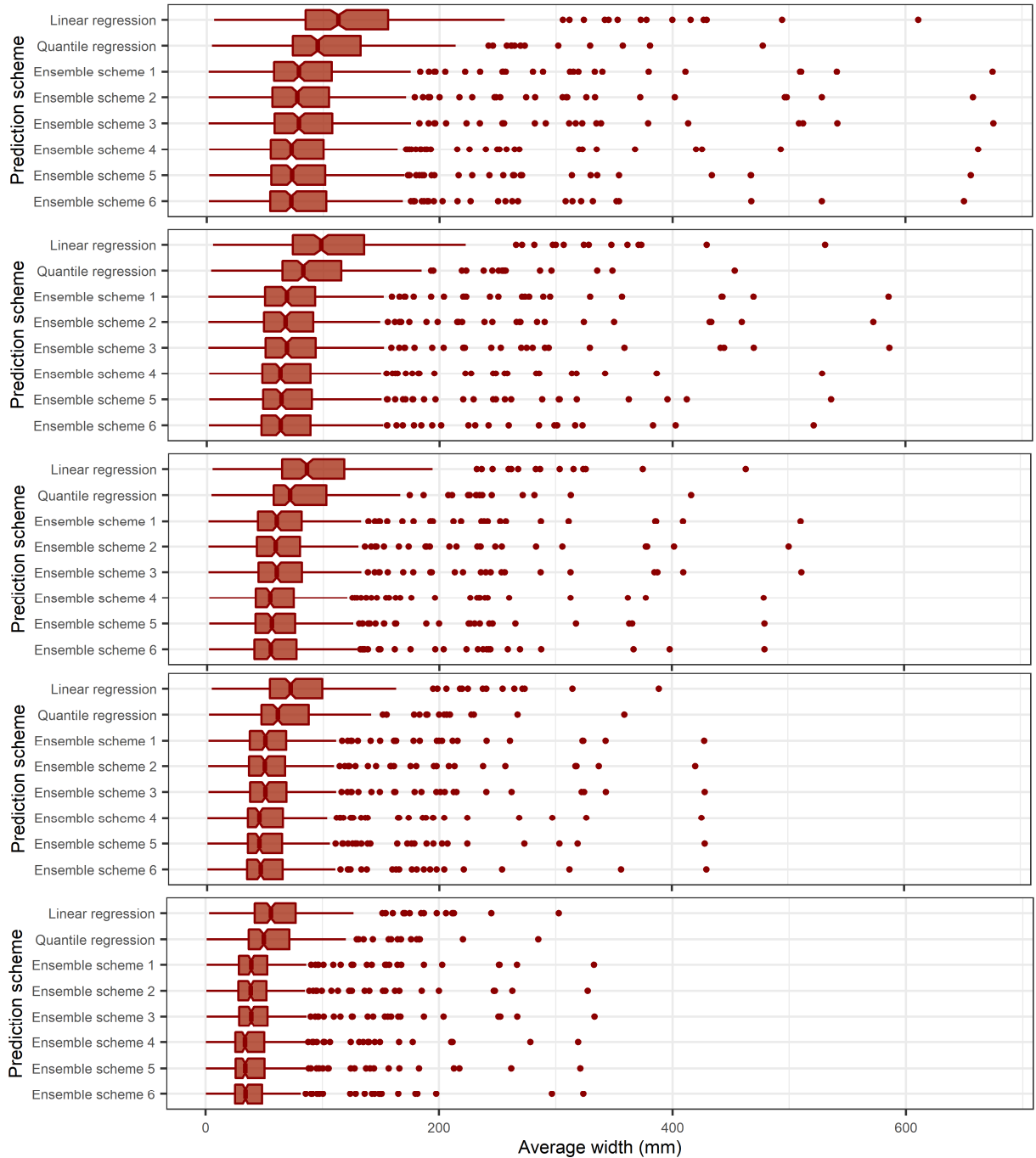

Figure 5. Average widths computed for the 99%, 97.5%, 95%, 90% and 80% prediction intervals (from top to bottom) delivered by the compared schemes for the period $T_3$ (years 1975–1999). Each boxplot summarizes 270 values.

The above-outlined information is objectively summarized in the average interval scores. The latter are collectively presented in Figure 6. The main information extracted from this figure is that (a) ensemble schemes 1–3, as well as ensemble schemes 4–6, exhibit very close performance to each other, (b) each ensemble scheme exhibits a better overall performance than its corresponding basic scheme, and (c) ensemble schemes 1–3 perform better than the quantile regression scheme for the 90% and 80% prediction intervals. Observation (b) indicates that the herein adopted implementations of the working methodology have an advantage over the naïve implementations of the data-driven (or purely statistical) models. This advantage should be further investigated before any generalization is made; nevertheless, this additional investigation involving, for instance, utilization of more predictor variables, goes beyond the aim of the present study.

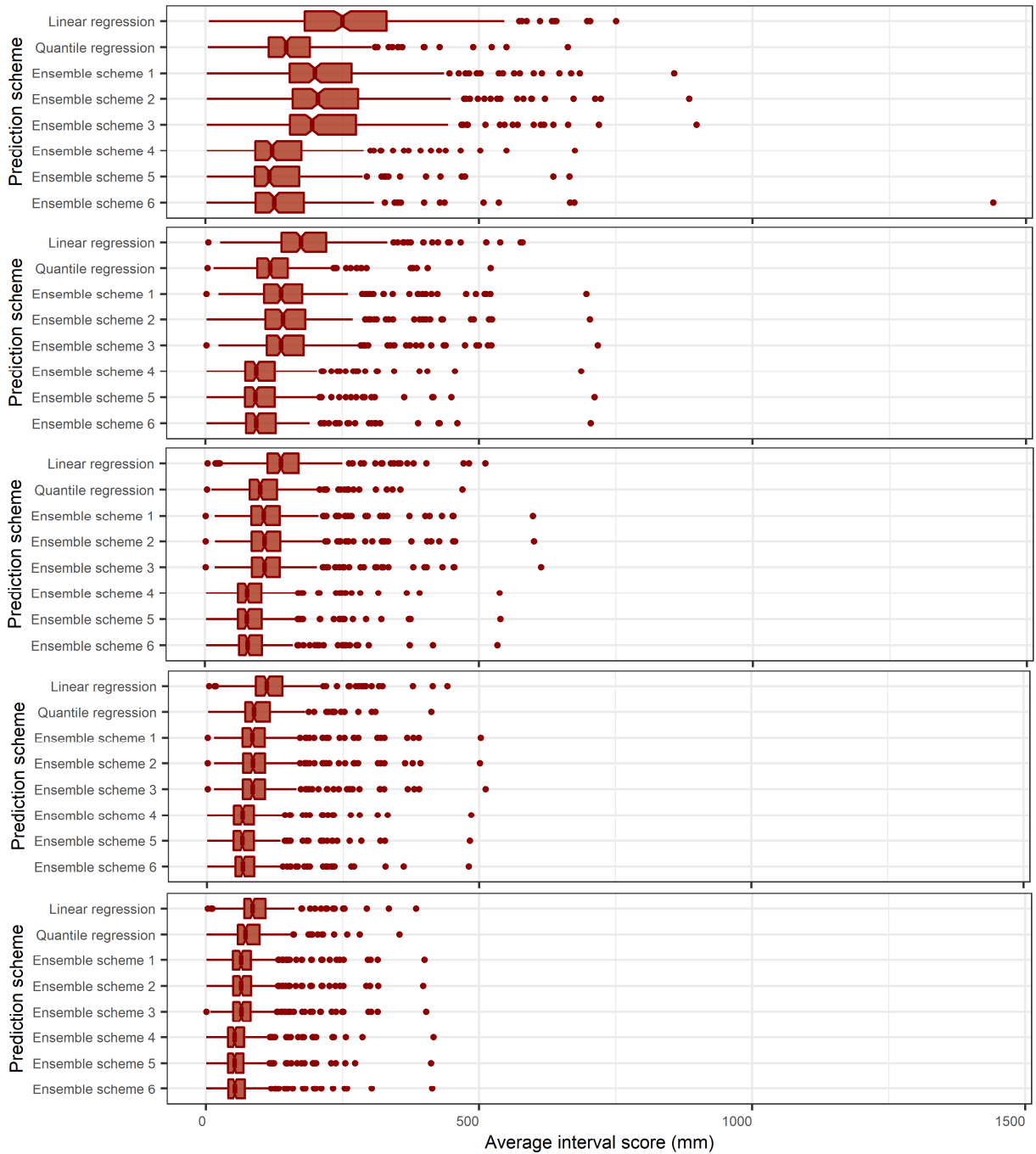

Figure 6. Average interval scores computed for the 99%, 97.5%, 95%, 90% and 80% prediction intervals (from top to bottom) delivered by the compared schemes for the period $T_3$ (years 1975–1999). Each boxplot summarizes 270 values.

We also note that both observations (a) and (b) are roughly expected already from the examination of Figures 4 and 5. By examining the aggregated average interval scores we additionally observe that the differences with respect to this metric are in average smaller for the inner prediction intervals than for the outer ones (as expected; see Section 2.3). Some small differences in the performance of ensembles schemes 1–3, favouring to a small extent ensemble schemes 1 and 3 over ensemble scheme 2, are mostly noticeable for the 99% and 97.5% prediction intervals. Similarly, ensemble scheme 5 seems to perform slightly better than ensemble scheme 4 for the same prediction intervals. It is also more effective than ensemble scheme 6 for all five prediction intervals.

To further inspect all differences, both the smaller and larger ones, in terms of rankings, the latter resulted for each catchment and for each examined prediction interval according to the computed average interval scores, we present Figures 7 and 8. The maps displayed in the former figure correspond to the upper side-by-side boxplots displayed on Figure 6, while allowing the examination of the rankings resulted both per catchment and per prediction scheme. From these maps we perceive that ensemble scheme 5 is ranked in a better average position than the remaining prediction schemes for the 99% prediction intervals, closely followed by ensemble schemes 4 and 6. Moreover, the quantile regression scheme is mostly ranked above the linear regression scheme and ensemble schemes 1–3. These schemes are mostly ranked in the last four positions. Importantly, there is not a fixed ranking position for any of the prediction schemes across the various catchments, while there are also some few catchments in which the four less competitive ones perform better than some the remaining. The quantile regression scheme is also ranked in the first three positions for a sufficient number of catchments. These latter observations provide us with a good reason to always perform large-scale benchmark experiments instead of (or alongside with) case studies.

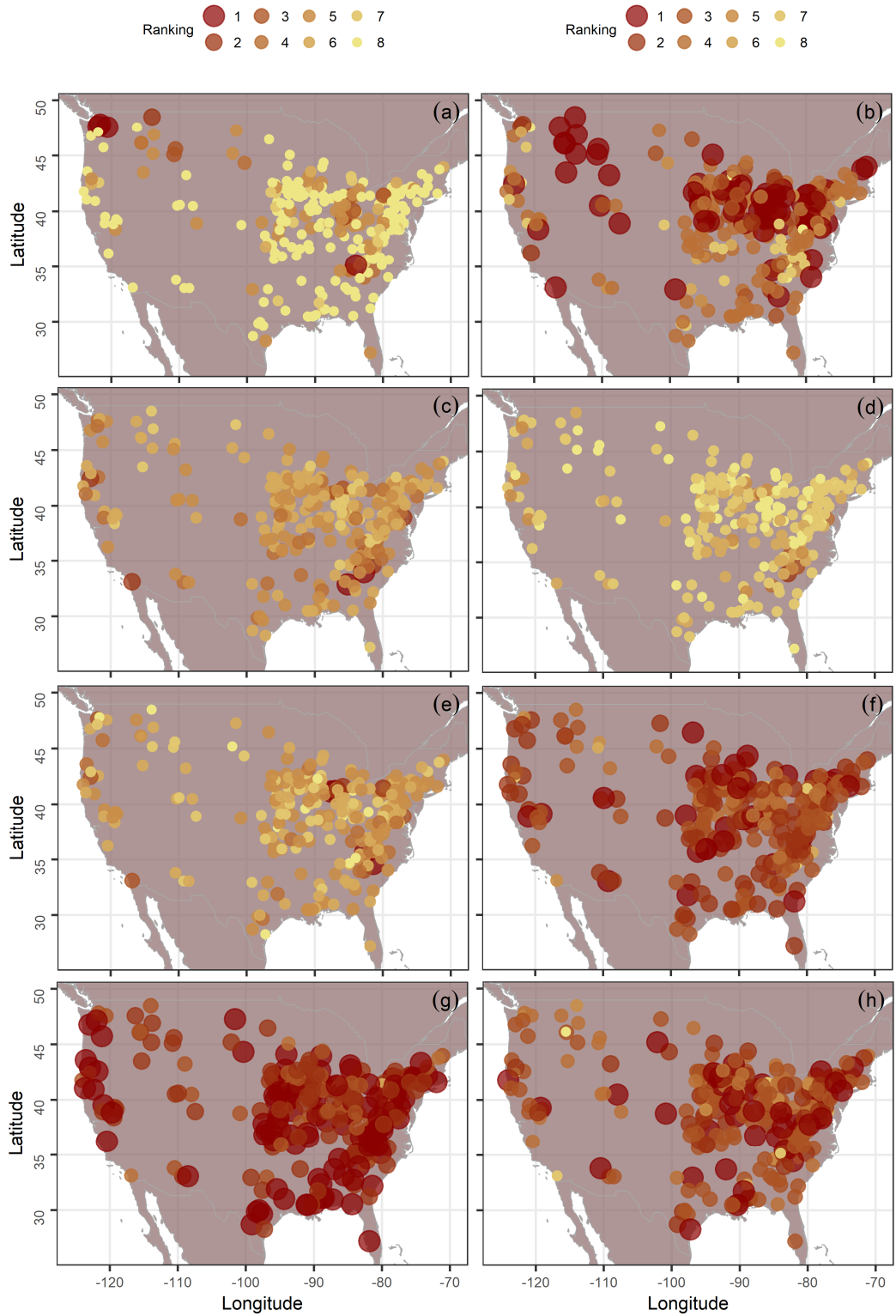

Figure 7. Rankings of (a) linear regression, (b) quantile regression and ensemble schemes (c–h) 1–6 according to the average interval scores computed for the 99% prediction intervals delivered for the period $T_3$ (years 1975–1999). The prediction schemes are ranked from best (1st) to worst (8th).

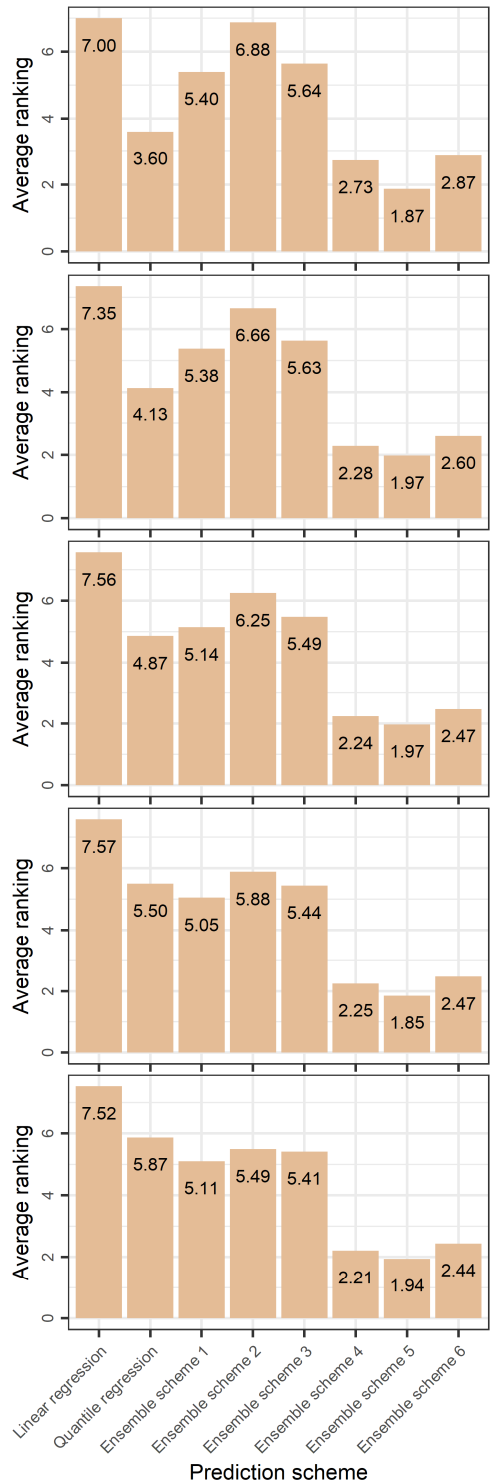

Figure 8. Average rankings of the prediction schemes according to the average interval scores computed for the 99%, 97.5%, 95%, 90% and 80% prediction intervals (from top to bottom) delivered by the compared schemes for the period $T_3$ (years 1975–1999). The prediction schemes are ranked from best (1st) to worst (8th). Each bar summarizes 270 values.

Overall, the image depicted in Figure 7 is rather neat when contrasted with its corresponding image in a similar visualization by Tyralis and Papacharalampous (2018); see Figure 4 therein. The latter study presents a large-scale comparison of point prediction methods that are equivalent to each other in a long run; therefore, no pattern is observed in their performance when the latter is depicted in maps. The pattern clearly observed in Figure 7, favouring the quantile regression model over the linear regression one, is due to the suitability of the former algorithm for modelling heteroscedasticity. Thus, it is our knowledge on the examined problem and the difference in the appropriateness of the adopted methodologies that created this pattern rather than anything else.

As emphasized in Papacharalampous et al. (2019a), only our knowledge on the system could make a tangible difference in (predictive) modelling in a long run. In fact, the homoscedasticity assumption is known to be inefficient when made during the probabilistic modelling of hydrological variables, such as the monthly river discharge variables that are of interest herein (see the comments, e.g., in Schoups and Vrugt 2010; Montanari and Koutsoyiannis 2012; Evin et al. 2013, 2014). Therefore, more flexible algorithms not assuming homoscedasticity are a reasonable choice to be made in such cases, while the same algorithms do not offer anything in comparison with less flexible algorithms in modelling cases where the homoscedasticity assumption is reasonable; see also Papacharalampous et al. (2019b), in particular the results displayed in Tables 4 and 5 for an illustration-justification of this fact.

The greatest part of the ranking-related information extracted from Figure 7 applies as well to the remaining prediction intervals, while a summary of this information for the 99%, 97.5%, 95%, 90% and 80% prediction intervals, presented in Figure 8, provides additional observations. The latter effectively complement those obtained from Figure 6. In fact, for all prediction intervals ensemble scheme 5 exhibits the best average-case ranking, closely followed by ensemble schemes 4 and 6. Moreover, the quantile regression scheme exhibits a significantly better (comparable) average-case ranking than (with) ensemble schemes 1–3 for the 99% and 97.5% (95%, 90% and 80%) prediction intervals, while the linear regression scheme is the worst performing in terms of average rankings, as it could be expected already from Figure 6.

To obtain a more faithful image of the gain or loss in performance when using each prediction scheme over the remaining ones, in Figure 9 we present the side-by-side boxplots of the relative improvements in terms of average interval score with respect to the linear regression scheme, while in Figure 10 we present the respective information using the quantile regression scheme as a reference. The closeness in the performance of ensembles schemes 1–3 is also perceivable by the examination of these figures. The same applies to the closeness in the performance of ensemble schemes 4–6. Nevertheless, the small differences favouring ensemble schemes 1 and 3 over ensemble scheme 2, and ensemble scheme 5 over ensemble schemes 4 and 6 are also highlighted. Additionally, we observe that the differences in the relative performance of a specific prediction scheme can be large, while there are cases in which the ensemble schemes are (far) worse than their respective basic schemes. However, the long-run image clearly favours the former over the latter, as already expected from the preceding visualizations.

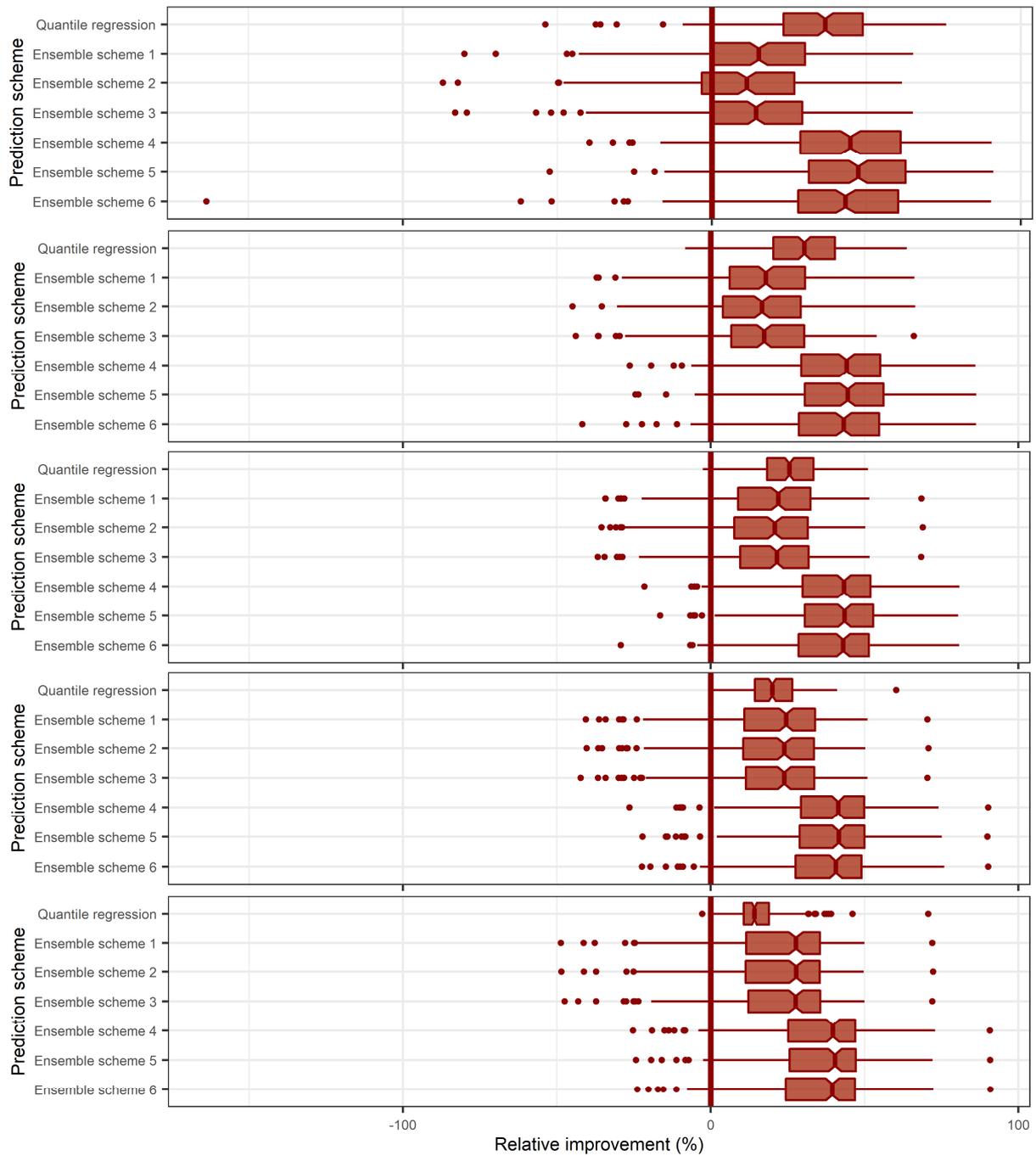

Figure 9. Relative improvements in terms of average interval score with respect to the linear regression scheme for the 99%, 97.5%, 95%, 90% and 80% prediction intervals (from top to bottom) delivered by the compared schemes for the period $T_3$ (years 1975–1999). Each boxplot summarizes 270 values. The reference values (zero values) are denoted with red thick vertical lines.

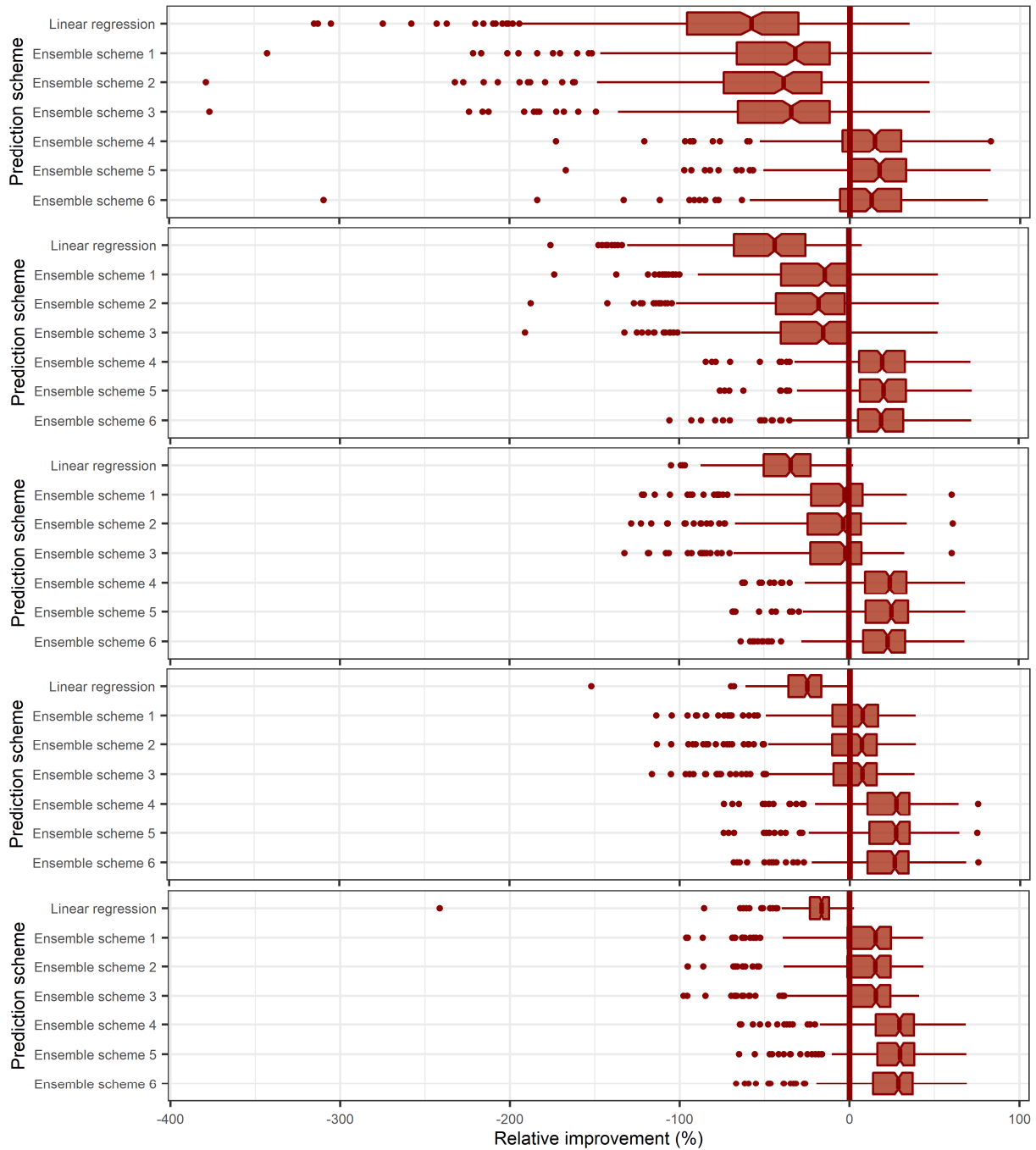

Figure 10. Relative improvements in terms of average interval score with respect to the quantile regression scheme for the 99%, 97.5%, 95%, 90% and 80% prediction intervals (from top to bottom) delivered by the compared schemes for the period $T_3$ (years 1975–1999). Each boxplot summarizes 270 values. The reference values (zero values) are denoted with red thick vertical lines.

We subsequently provide a numerical summary of the gain in performance when using specific schemes over others, as extracted from the real-world experiment of the study. In Figures 11 and 12 we present the average-case relative improvements in terms of average interval score with respect to the linear regression and the quantile regression schemes respectively. These two figures objectively summarize the information presented in Figures 9 and 10, while they are particularly useful in assessing how small the differences between ensemble schemes 1–3, as well as between ensembles schemes 4–6, are; see also Figures S.1 and S.2 of the supplementary material (see Appendix D) for inspecting these differences in terms of median relative improvements. For the former category of ensemble schemes, we observe that the difference in the average-case improvements is at maximum 3.65%. The latter difference is computed for ensemble schemes 1 and 2 for the 99% prediction intervals, while it is smoothened to 1.94%, 1.07%, 0.48% and 0.13% for the 97.5%, 95%, 90% and 80% prediction intervals respectively. The average relative improvements when using ensemble scheme 1 instead of ensemble scheme 2 are 4.24%, 2.39%, 1.36%, 0.63% and 0.18% for the obtained 99%, 97.5%, 95%, 90% and 80% prediction intervals. The respective median improvements are 3.75%, 2.18%, 1.20%, 0.53% and 0.15%, while the cost in terms of computational time is about 12 min for all 270 catchments. Ensemble scheme 3 offers comparable profit in performance alongside with a 28-minute profit in terms of computational time compared to ensemble scheme 1.

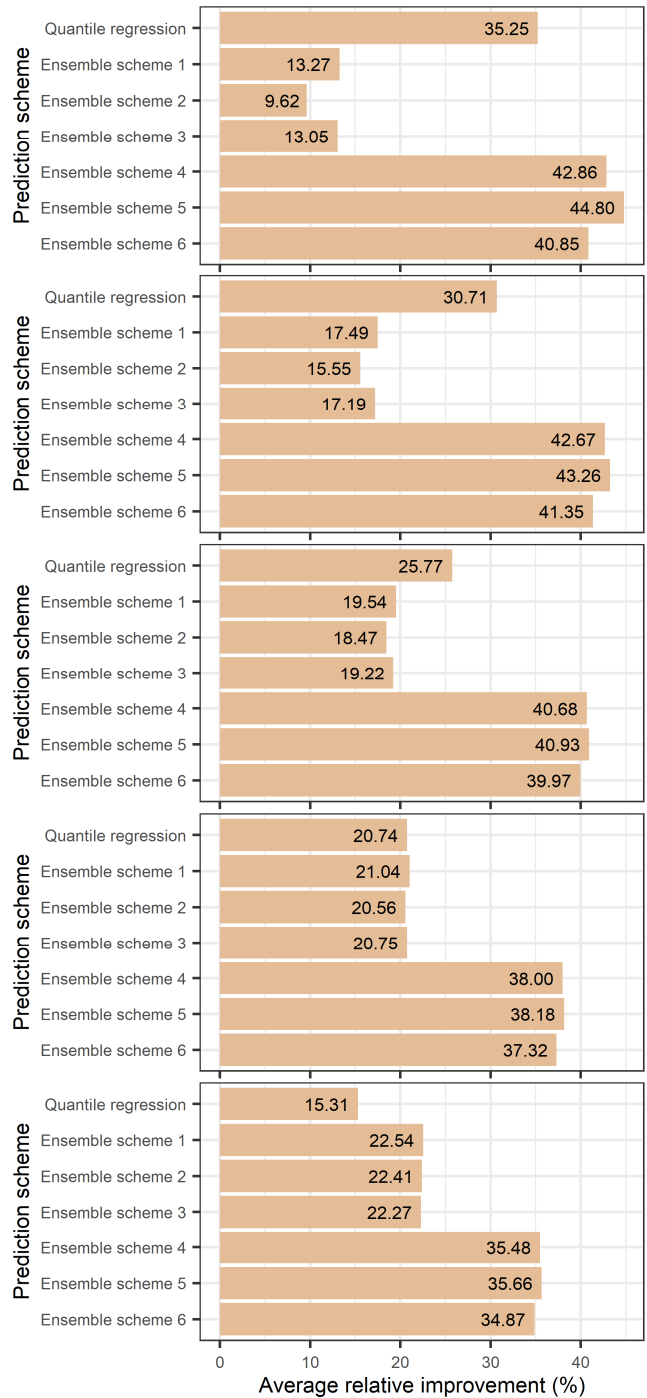

Figure 11. Average relative improvements in terms of average interval score with respect to the linear regression scheme for the 99%, 97.5%, 95%, 90% and 80% prediction intervals (from top to bottom) delivered by the compared schemes for the period $T_3$ (years 1975–1999). Each bar summarizes 270 values.

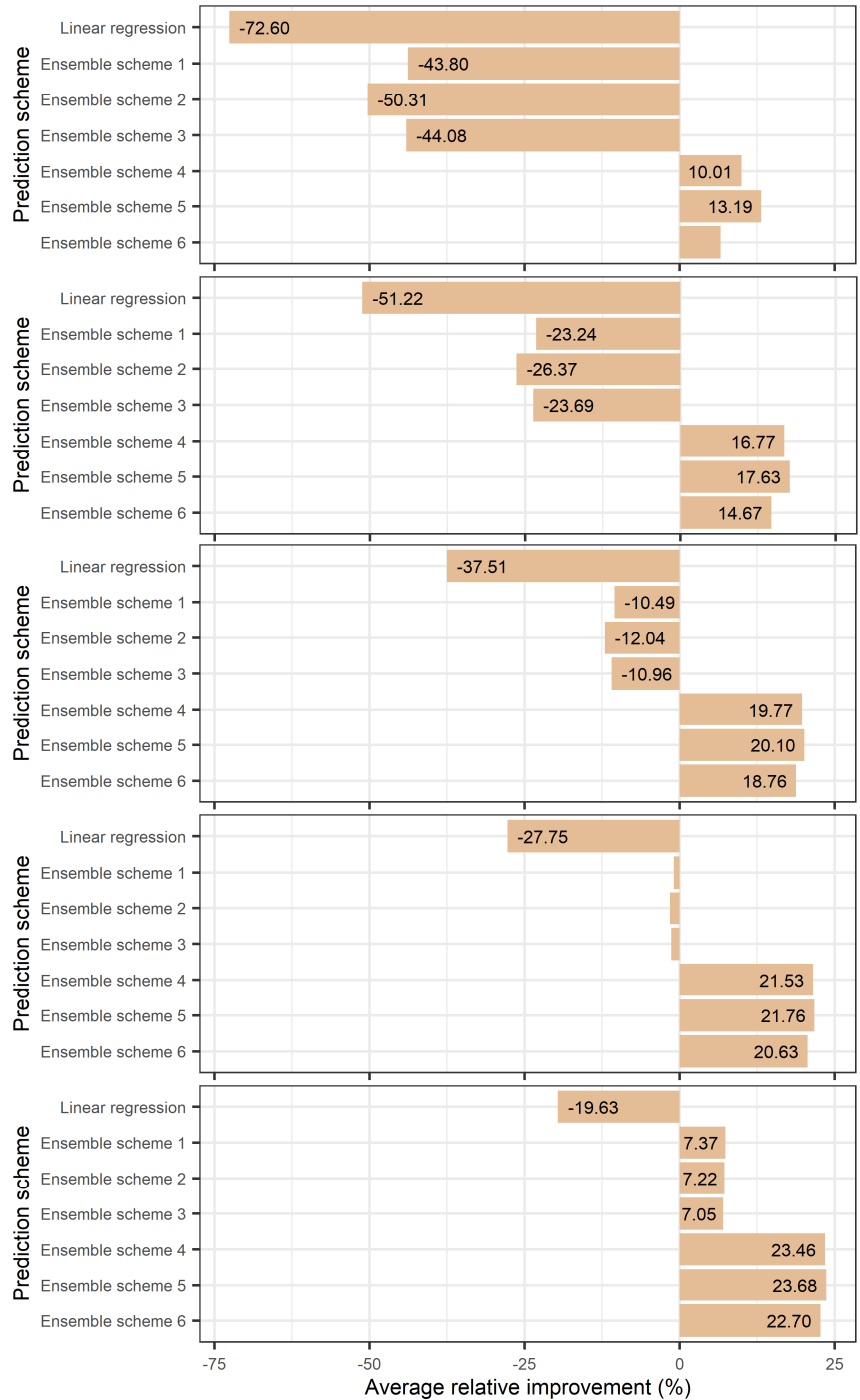

Figure 12. Average relative improvements in terms of average interval score with respect to the quantile regression scheme for the 99%, 97.5%, 95%, 90% and 80% prediction intervals (from top to bottom) delivered by the compared schemes for the period $T_3$ (years 1975–1999). Each bar summarizes 270 values.

Moreover, the mean (median) profit when using ensemble scheme 5 instead of ensemble scheme 4 is found to be 3.09%, 0.99%, 0.48%, 0.34% and 0.25% (2.07%, 0.54%, 0.32%, 0.27% and 0.18%) for the 99%, 97.5%, 95%, 90% and 80% prediction intervals respectively, while the concomitant cost in terms of computational time is about 36 min. The respective profit when using ensemble scheme 6 over ensemble scheme 4 is about 12 min. Nonetheless, the use of the latter scheme instead of the former scheme offers an average (median) relative improvement equal to 2.23%, 1.77%, 1.11%, 1.00% and 0.85% (0.31%, 0.47%, 0.24%, 0.28% and 0.31%) for the 99%, 97.5%, 95%, 90% and 80% prediction intervals respectively. Moreover, the respective average (median) relative improvements provided by ensemble scheme 5 with respect to ensemble scheme 6 are 5.46%, 2.74%, 1.60%, 1.36%, 1.10% (3.39%, 1.44%, 0.73%, 0.57%, 0.45%). The gain in performance from the incorporation into the working methodology of the quantile regression model instead of the linear regression model can be summarized by the average-case (median) relative improvements in terms of average interval score provided when using ensemble scheme 5 instead of ensemble scheme 1. These are 37.00%, 31.62%, 26.82%, 22.10% and 17.22% (37.97%, 31.32%, 25.85%, 20.95% and 15.84%) for the 99%, 97.5%, 95%, 90% and 80% prediction intervals respectively.

## 3.2 Addressing aims 4–5 of the study

Two key properties of the working methodology, as identified in Papacharalampous et al. (2019b) based on the seminal work by Lichtendahl et al. (2013, Section 5), are its larger robustness in performance compared to basic two-stage post-processing methodologies and its ability to harness the wisdom of the crowd, both stemming from the concept of prediction averaging. These properties can also be considered as the result of an optimal exploitation of the possibilities offered by the MK blueprint methodology. The demonstration of these properties has only been made so far within toy examples, while it is still pending for rainfall-runoff problems. This section is devoted to empirically proving these two properties of the working methodology using the results of the herein conducted real-world experiment, i.e., to addressing aims 4–5 of the study. These aims are of particular importance in justifying the conceptualization and rationale behind the working methodology.

In Figure 13 we present the relative improvements when using the output of ensemble scheme 5, i.e., the average of 600 quantile predictions, instead of separately using each of them (i.e., the relative improvements {$RI_{OUT,IN_i}$, $i$ = 1, …, 600}, defined with Equation 6, for ensemble scheme 5), computed for all catchments and for all prediction intervals. We observe that these relative improvements are approximately symmetric around zero, in average slightly higher than zero. Specifically, the average relative improvements corresponding to Figure 13 are found to be equal to 0.82%, 0.83%, 0.74%, 0.70% and 0.71% for the 99%, 97.5%, 95%, 90% and 80% prediction intervals respectively (see Table S.1). The interpretation of this outcome is straightforward, while indicating an advantage in terms of robustness of the working methodology over basic two-stage post-processing methodologies using a single probabilistic prediction.

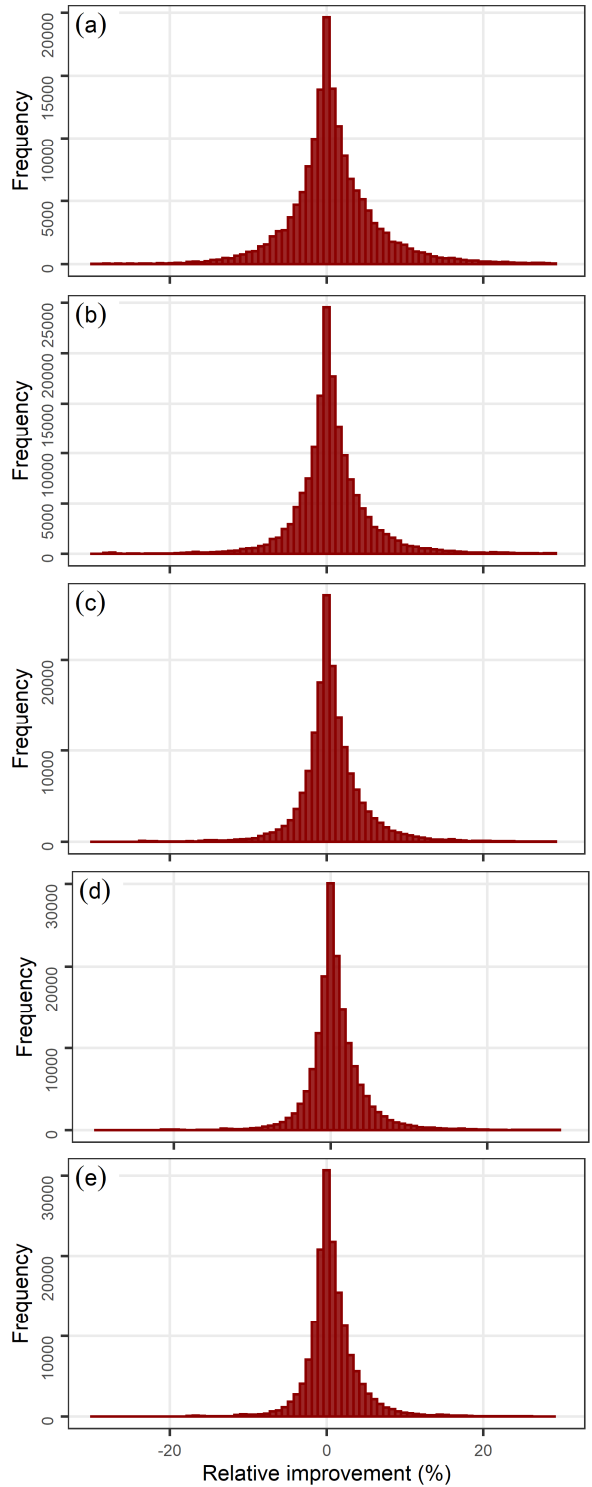

Figure 13. Relative improvements $\{RI_{OUT,IN_i}, i = 1, ..., 600\}$ (defined with Equation 6) for ensemble scheme 5. The relative improvements are computed for all catchments, and for the (a) 99%, (b) 97.5%, (c) 95%, (d) 90% and (e) 80% prediction intervals obtained for the period $T_3$ (years 1975–1999). The horizontal axis has been truncated at −30% and 30%. Each histogram summarizes 270 × 600 = 162 000 values.

In fact, while approximately half of the probabilistic predictions score better (or worse) than the finally delivered by the working methodology probabilistic prediction, there is no way to know in advance which hydrological model's parameters will lead in better average interval score in the period $T_3$. While this lack of knowledge could significantly affect (in terms of performance) the delivered probabilistic prediction for a basic two-stage post-processing methodology, this effect is largely reduced by the working methodology.

Moreover, by comparing the degree of spread in the five histograms displayed in Figure 13, we also perceive that the degree of the offered stabilization in performance seems to become larger as we move from the inner prediction intervals to the more outer ones. Nevertheless, even for the 80% prediction intervals the provided stabilization is significant.

Furthermore, in Figure 14 we present the relative differences between the average interval score of the output of ensemble scheme 5 and the average of the average interval scores of each of the combined (for obtaining this output) individual predictions, the latter used as reference for the former (i.e., the relative differences $RD_{OUT,AAIS_{IN}}$, defined with Equation 8, for ensemble scheme 5), computed for all catchments and for all prediction intervals. Importantly, all computed relative differences are positive (or approximately zero) with no exception; therefore, the average of quantile predictions scores no worse than the average score of the combined individual predictions, i.e., the working methodology harnesses the wisdom of the crowd in terms of average interval score when applied for solving monthly rainfall-runoff problems (see also Lichtendahl et al. 2013, Section 5). The average relative differences corresponding to Figure 14 are 1.30%, 1.12%, 0.94%, 0.85% and 0.84% for the 99%, 97.5%, 95%, 90% and 80% prediction intervals respectively (see Table S.2).

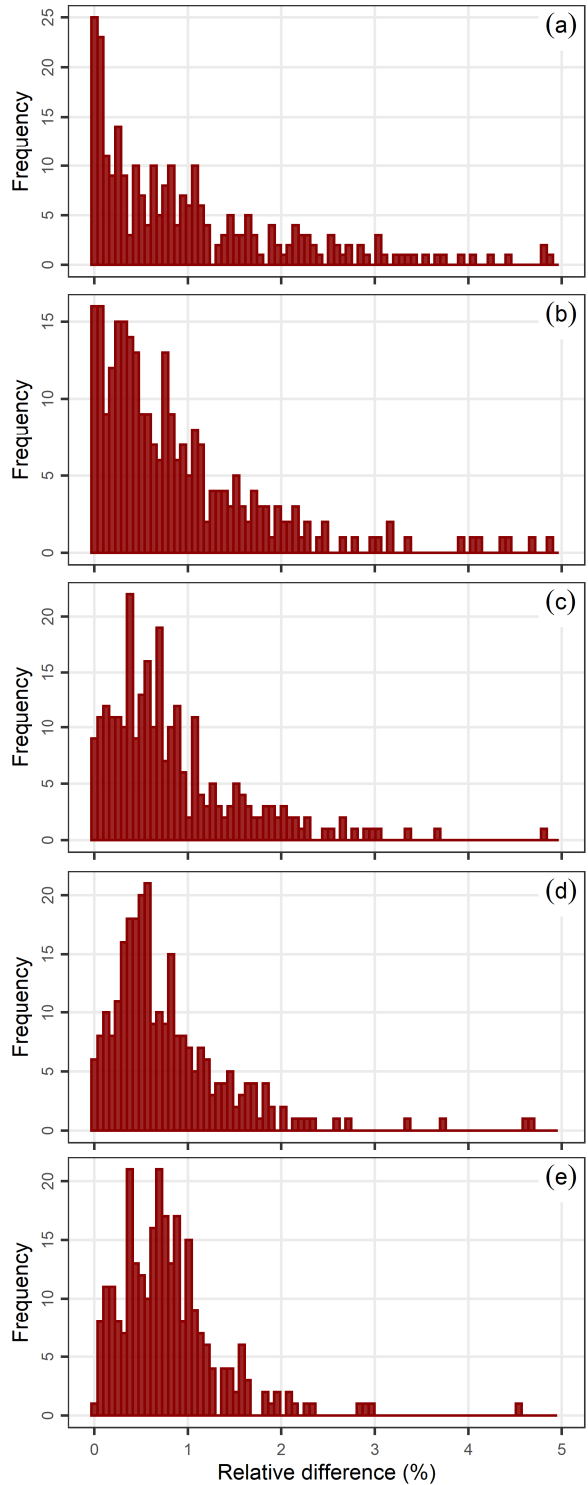

Figure 14. Relative differences $RD_{OUT,AAIS_{IN}}$ (defined with Equation 8) for ensemble scheme 5. The relative differences are computed for all catchments, and for the (a) 99%, (b) 97.5%, (c) 95%, (d) 90% and (e) 80% prediction intervals obtained for the period $T_3$ (years 1975–1999). The horizontal axis has been truncated at 5%. Each histogram summarizes 270 values.

Analogous observations are extracted from analogous investigations for all remaining ensemble schemes (see Figures S.3–S.12 and Tables S.1–S.2 of the supplementary material). In summary, the relative improvements when using the output of an ensemble scheme, i.e., the average of 600 quantile predictions, instead of separately using each of these predictions range from −327.10% to 91.42%. The average of these relative improvements ranges between 0.13% and 1.13%. Similarly, the average relative differences favouring the average interval score computed for the output of an ensemble scheme over the average of the average interval scores computed for each of the combined (for obtaining this output) individual predictions range between 0.19% and 1.83%. The average relative improvement (difference) is in general larger for the outer prediction intervals than for the inner ones, while its magnitude also depends on the ensemble scheme.

As also emphasized in Papacharalampous et al. (2019b), the overall trade-off to be considered when someone has to choose between the working methodology and a basic two-stage post-processing methodology allowing the utilization of the same type of flexible error models (see e.g., López López et al. 2014; Dogulu et al. 2015; Papacharalampous et al. 2019d) is the one between (a) the larger robustness in performance offered by the former methodology (demonstrated in Figures 13, S.3, S.5, S.7, S.9 and S.11, and Table S.1) and the ability of this methodology to harness the wisdom of the crowd (empirically proven based on Figures 14, S.4, S.6, S.8, S.10 and S.12, and Table S.2), and (b) the significantly less computational requirements of the latter methodologies.

## 4. Concluding remarks

We have validated the probabilistic hydrological modelling methodology proposed in Papacharalampous et al. (2019b). This methodology adopts key concepts from the ensemble post-processing methodology by Montanari and Koutsoyiannis (2012), while also relying on the concept of probabilistic prediction combination from the forecasting field. It applies a single hydrological model using a large number of different parameter values to generate the same number of "sister predictions". The parameters of the hydrological model can be obtained by using either Bayesian calibration schemes or informal calibration schemes (see the related investigations in Appendix E). Therefore, this methodology does not have any particular relationship with Bayesian methods by construction, as it also applies to its precursor. A statistical learning (or machine learning) regression model that is suitable for predicting quantiles (see e.g., the models exploited in Papacharalampous et al. 2019d) is then used to obtain information about the hydrological model's error. This information is used to convert the sister predictions into probabilistic predictions, which are finally combined in simple fashion to obtain the output probabilistic predictions. The assessed methodology is subdivided into three alternative variants, which differ only in the training of the regression model.

We have conducted a large-sample real-world experiment at monthly timescale, set up using complete 50-year daily information for 270 catchments in the United States. Aiming to increase the understanding in probabilistic hydrological modelling, we have insisted on interpretability and benchmarking within all conducted tests. We have used the parsimonious GR2M hydrological model and two (largely) interpretable regression models, specifically the linear regression and the quantile regression ones, to implement six ensemble schemes, all of them based on the assessed methodology. Those ensemble schemes implemented using the linear model (three in number) have been used as benchmarks for the remaining schemes (also three in number). Those ensemble schemes using the same regression model rely on different variants of the assessed methodology. The performance of the ensemble schemes has been assessed by computing the coverage probabilities, average widths and average interval scores of the obtained interval predictions, and by also benchmarking their results using naïve probabilistic data-driven models.

The obtained numerical results (metric values computed for 4 870 800 interval predictions) suggest the usefulness of the assessed methodology in obtaining probabilistic predictions of hydrological quantities. The best-performing variant, offering a mean relative improvement up to 5.46% with respect to its alternative variants, when implemented using the quantile regression model, is variant 2. This variant trains the regression model on a single large dataset formed by using information from all sister predictions. The average-case relevant improvements when using the quantile regression model instead of the linear regression one range up to about 37% in terms of average interval score. This latter numerical result should be appraised on the basis that only the former of these models can model heteroscedasticity. The homoscedasticity assumption is often made in the literature when modelling the hydrological model's error.

Finally, we have demonstrated the increased robustness of the assessed methodology with respect to the combined (by this methodology) individual predictors and, by extension, to basic two-stage post-processing methodologies. The ability to "harness the wisdom of the crowd" has also been empirically proven. The quantile predictions obtained by all ensemble predictors are found to score no worse –usually better– than the average of the individual scores of the combined individual predictions in terms of average interval score. This outcome is in line with demonstrations for stylized cases by Lichtendahl et al. (2013). The computed relative differences favour the former quantity over the latter up to about 37%, while their mean values range between 0.19% and 1.83%, depending both on the prediction interval and the variant of the assessed methodology. For the best-performing ensemble scheme the respective average relative differences are around 1%. Overall, the robustness and the ability to harness the wisdom of the crowd are identified as two key properties of the working methodology.

**Appendix A    Background methodological considerations**

In this appendix, we summarize in terms of advantages and disadvantages some technical and theoretical considerations that currently guide the selection between Bayesian and two-stage post-processing methodologies for uncertainty assessment in the field. This summary is mainly presented through Tables A.1 and A.2. Moreover, in Table A.3 we list the advantages and disadvantages offered by statistical learning (or machine learning) quantile regression algorithms, since these algorithms serve as error models within the working methodology.

Table A.1. Advantages and disadvantages of Bayesian hydrological post-processing methodologies (see also Evin et al. 2014). These post-processing methodologies jointly infer (within a Bayesian framework) the parameters of the hydrological and error models by using the entire historical dataset.

| | |
|---|---|
| Advantages | o If their assumptions are proper, they produce optimal probabilistic predictions by theory. This could be possible in principle, since the hydrological literature presents generalized findings on the distributions of hydrological variables with increasing frequency and reliability.<br>o They can largely facilitate interpretability in modelling, since they allow the inspection of the impact of their assumptions on both parameter and predictive uncertainty.<br>o Their performance depends less on the length of the historical dataset than the performance of two-stage post-processing methodologies (see Table A.2), since their fitting does not require sample splitting. |
| Disadvantages | o Their predictive performance largely depends on the appropriateness of their assumptions.<br>o They might get over-parameterized in an effort to ensure the adoption of proper assumptions.<br>o Their use is accompanied by computational limitations. |

Table A.2. Advantages and disadvantages of two-stage hydrological post-processing methodologies (see also Evin et al. 2014; Papacharalampous et al. 2019d, Section 5.2.2). These post-processing methodologies estimate their error models conditional on the predictions provided by their hydrological models. The latter have been calibrated by using an independent segment of the historical dataset.

| | |
|---|---|
| Advantages | o They can be nearly assumption-free (i.e., their performance does not necessarily depend on the appropriateness of assumptions) when implemented with flexible machine learning quantile regression algorithms as error models. The advantages of these algorithms are listed independently in Table A.3.<br>o Computational requirements and limitations are mostly few in their case. Therefore, their automation and application to big datasets is feasible. This is one of the main reasons why two-stage hydrological post-processing is popular in forecasting applications. This popularity is emphasized e.g., by Evin et al. (2014).<br>o In light of the two points above, their performance can be maximized by adopting algorithmic strategies and well-established guidelines from the machine learning literature (see e.g., the experiment presented herein). The role of big datasets for achieving optimal modelling solutions under this new-era approach is emphasized e.g., in Tyralis et al. (2019b). |
| Disadvantages | o They largely lack interpretability by perception. Interactions between the hydrological model parameters and the trained version of the error model are ignored; therefore, their hydrological model parameter estimates are only auxiliary to predictive uncertainty quantification and cannot be used in any case for understanding parameter uncertainty.<br>o Their performance depends more on the length of the historical dataset than the performance of Bayesian post-processing methodologies (see Table A.1), since their fitting requires sample splitting.<br>o The adoption of flexible machine learning quantile regression algorithms as error models has an additional cost in terms of interpretability and further increases the large-sample requirements (see the disadvantages of Table A.3). These requirements are revealed and discussed e.g., in Papacharalampous et al. (2019b, Appendix D). |

Table A.3. Advantages and disadvantages of statistical learning (or machine learning) quantile regression algorithms (see also Waldmann 2018; Papacharalampous et al. 2019d, Sections 2.3.1, 5.2.2). Quantile regression algorithms issue quantile predictions instead of PDF predictions.

| | |
|---|---|
| Advantages | o They are ideal when the conditional distribution of the dependent variable is not known or is hard to deduce. |
| | o They model heteroscedasticity by perception and construction. |
| | o In light of the above point, they are also straightforward to apply, as they do not need to be fitted separately for each season (or month), in contrast to distribution-based modelling approaches (e.g., conditional-distribution models). |
| | o They are robust with respect to outliers in the observations of the dependent variable. |
| | o They are available in open source and mostly optimally programmed. |
| Disadvantages | o They are trained separately for each quantile probability; therefore, the more the quantiles (or prediction intervals) we are interested in issuing, the more computationally costly the training process. |
| | o Quantile crossing is possible. |
| | o Parameter estimation is harder than in standard regression. |
| | o Their performance depends to some extent on the sample size. |
| | o They lack interpretability. Only their linear variant, i.e., the quantile regression model implemented herein, offers interpretability to some extent. |

## Appendix B    Statistical software information

The analyses and visualizations have been performed in `R` Programming Language (R Core Team 2019). We have used the following contributed `R` packages: `airGR` (Coron et al. 2017, 2019), `bestNormalize` (Peterson 2017, 2019), `coda` (Plummer et al. 2006; 2019), `data.table` (Dowle and Srinivasan 2019), `devtools` (Wickham et al. 2019c), `dplyr` (Wickham et al. 2019b), `FME` (Soetaert and Petzoldt 2010, 2016), `gdata` (Warnes et al. 2017), `ggplot2` (Wickham 2016a; Wickham et al. 2019a), `ggridges` (Wilke 2018), `hddtools` (Vitolo 2017, 2018), `knitr` (Xie 2014, 2015, 2019), `maps` (Brownrigg et al. 2018), `matrixStats` (Bengtsson 2018), `plyr` (Wickham 2011, 2016b), `quantreg` (Koenker 2019), `readr` (Wickham et al. 2018), `reshape` (Wickham 2007, 2018), `rmarkdown` (Allaire et al. 2019), `tidyr` (Wickham and Henry 2019) and `zoo` (Zeileis and Grothendieck 2005; Zeileis et al. 2019). We have also followed procedures described in the contributed vignettes of the `airGR R` package (https://cran.r-project.org/web/packages/airGR/vignettes).

## Appendix C    Working methodology

This appendix is largely adapted from Papacharalampous et al. (2019b). It aims at summarizing the working methodology. For this summary, we first define the time period $T = \{1, ..., (n_1+n_2+n_3)\}$, and its three distinct sub-periods $T_1 = \{1, ..., n_1\}$, $T_2 = \{(n_1+1), ..., (n_1+n_2)\}$ and $T_3 = \{(n_1+n_2+1), ..., (n_1+n_2+n_3)\}$. We also define the sister model realizations as variants of a single hydrological model, each using different parameter values. The latter are obtained by calibrating the hydrological model in the period $T_1$. The calibration could be made by using either Bayesian schemes (e.g., Markov Chain Monte Carlo simulation sampling; see e.g., the procedures described in Section 2.2.4) or informal calibration schemes (see e.g., the procedures described in Appendix E). Let us assume that we obtain $m$ sister model realizations, where $m$ is adequately large. Each sister model realization is then applied in the period $\{T_2, T_3\}$. The $m$ resulted sister predictions also extend in the period $\{T_2, T_3\}$. We subsequently compute the sister model realizations' errors in the period $T_2$ by using the sister predictions alongside with their corresponding target values.

Information about the sister model realizations' error is then obtained by training a statistical learning regression model that is suitable for predicting quantiles (hereafter referred to as "error model"; see e.g., the error models exploited in Papacharalampous et al. 2019d) in the period $T_2$. In particular, we regress the sister model realizations' error at time $t$ (response variable) on selected predictor variables (e.g., the sister prediction at time $t$). For each sister prediction extending in the period $T_3$, we (a) predict a set of quantiles (with selected probabilities) of the sister model realization's errors using the information obtained at the preceding step, and (b) transform these predictive quantiles to auxiliary predictive quantiles of the hydrological process of interest (by subtracting them from their corresponding sister prediction). Finally, at each time $t \in T_3$ we group the auxiliary predictive quantiles of the hydrological process of interest based on their corresponding probability (e.g., probability 0.95) to average them over each group. The resulted time series are the output quantile predictions.

The basic steps adopted within the working methodology are also summarized in Figure B.1.

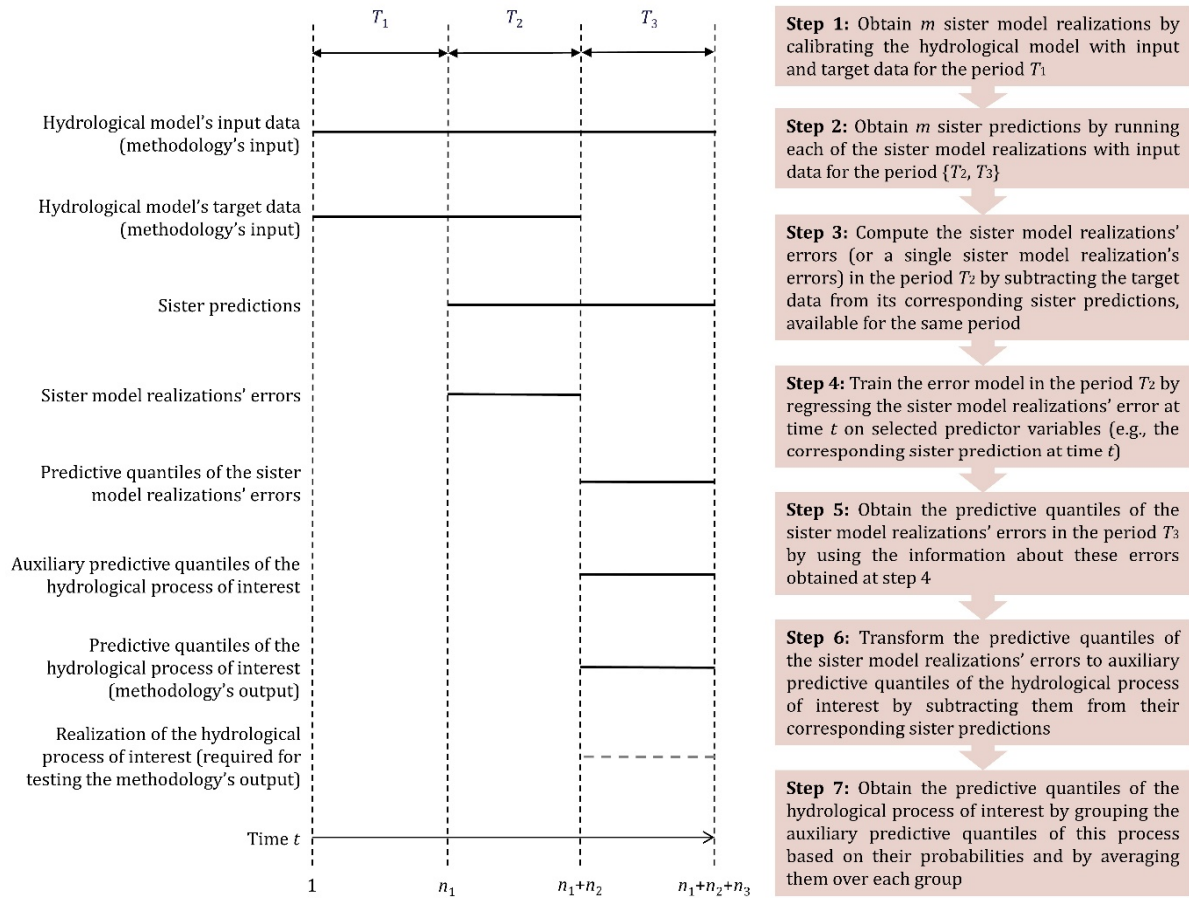

Figure B.1. Schematic summarizing the working methodology (reproduced from Papacharalampous et al. 2019b). The sister model realizations are defined as variants of a single hydrological model, each using different parameter values. The latter can either be drawn from the respective simulated posterior distribution of model parameters or can be obtained by using informal calibration schemes. Each sister model realization is used for obtaining a single point prediction, referred to as "sister prediction". The number of sister model realizations $m$ should be adequately large. The realization of the hydrological process of interest, considered unknown at the time of the prediction, is denoted with a light grey dashed line.

The working methodology is subdivided into three alternative variants. These variants differ in the error model's training only. Specifically:

o   Variant 1 trains the error model $m$ times, each time on a different dataset formed by using a different sister prediction;

o   Variant 2 trains the error model on a single dataset formed by using all sister predictions;

o   Variant 3 also trains the regression model once; however, the training here is made on a dataset formed by using one randomly selected sister prediction.

We note that the three variants reduce to the same method in the case that a single point hydrological prediction is generated. In this case, the working methodology would fall into the category of basic two-stage post-processing methodologies using regression models.

**Appendix D    Supplementary material**

The supplementary material to this article is available in Papacharalampous et al. (2019c). This material includes Figures S.1–S.12, and Tables S.1 and S.2. The latter are extracted from the large-scale investigations presented in Section 3.

**Appendix E    Additional investigations**

To investigate the possibility of using informal calibration schemes instead of Bayesian schemes for obtaining a large number of hydrological model's parameters within the working methodology, in this appendix we repeat the large-sample experiment of the study (only for the ensemble schemes) by using different parameter values for the hydrological model. Specifically, for each catchment we retain the first 200 parameter values from each simulated chain (see Section 2.2.4) that have not converged to the posterior distribution of the parameters, instead of the last 200 values that were previously retained (for the application presented in Section 3). Hereafter, let us refer to the calibration scheme adopted for obtaining the parameters of the hydrological model in the original large-sample experiment of the study (presented in Section 3) and the calibration scheme that is adopted in this appendix as "Bayesian calibration scheme" and "informal calibration scheme" respectively. The remaining components of the ensemble schemes are retained as detailed in Section 2.2.

Once we have obtained the interval predictions, we compute their interval scores and the relative improvements provided in terms of average interval score by the informal calibration scheme with respect to the Bayesian calibration scheme, when both these schemes are exploited as components of ensemble schemes 1–6. The computations are made as detailed in Section 2.3, while the related information is presented in Figure D.1. We mainly observe that (a) the relative improvements can be either positive or negative, and (b) the results favour the Bayesian calibration scheme to some extent, mostly due to outliers. These outliers may become fewer with increasing the length of the period $T_2$. To objectively summarize the derived information, we also compute the mean and median relative improvements in terms of the same score. These are presented in Figures D.2 and D.3 respectively.

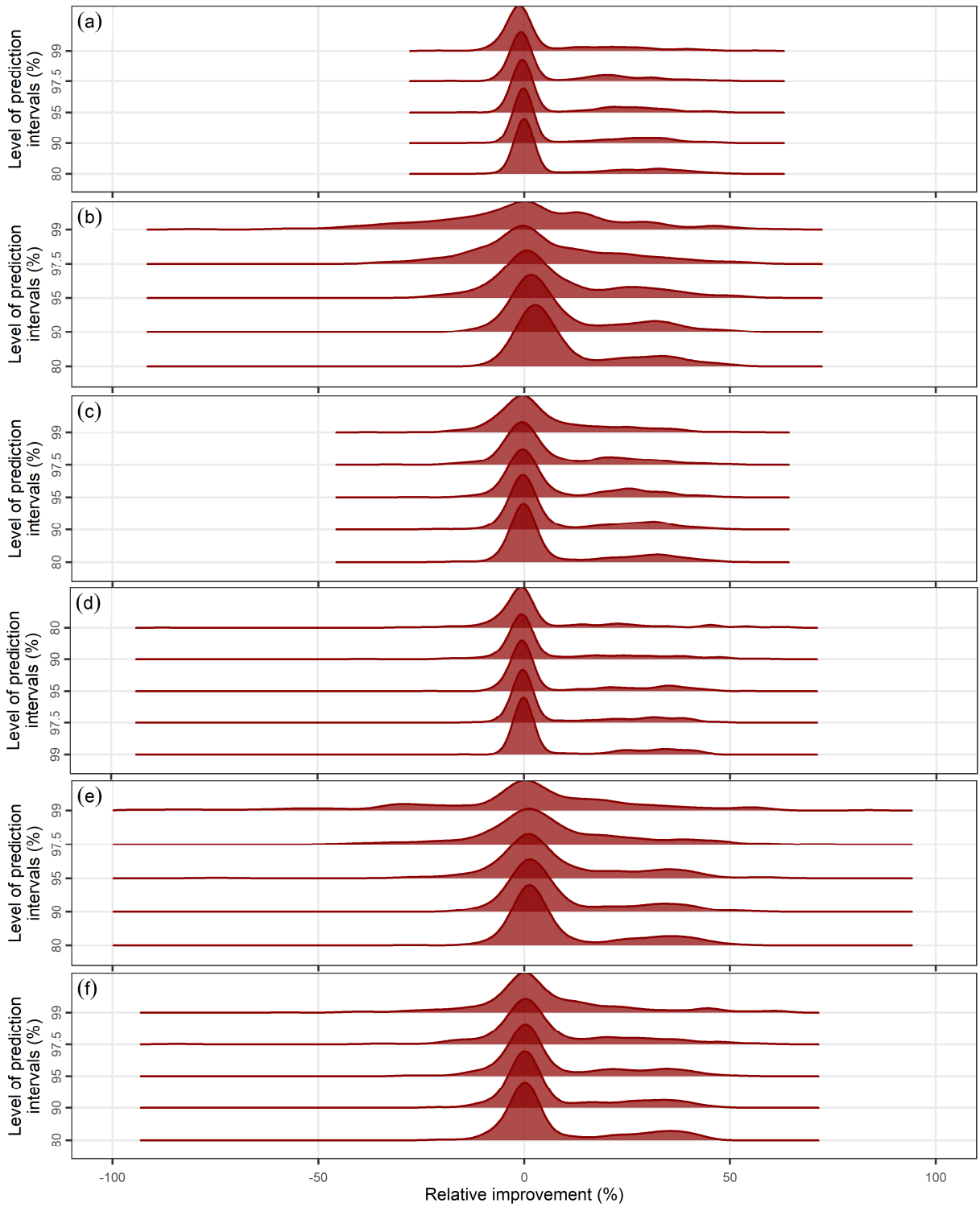

Figure D.1. Densities of the relative improvements in terms of average interval score provided by the Bayesian calibration scheme with respect to the informal calibration scheme, when both these schemes are used as components of (a–f) ensemble schemes 1–6. The latter are implemented with their remaining components and parameters set common. The horizontal axis has been truncated at −100% and 100%. Each density summarizes 270 values.

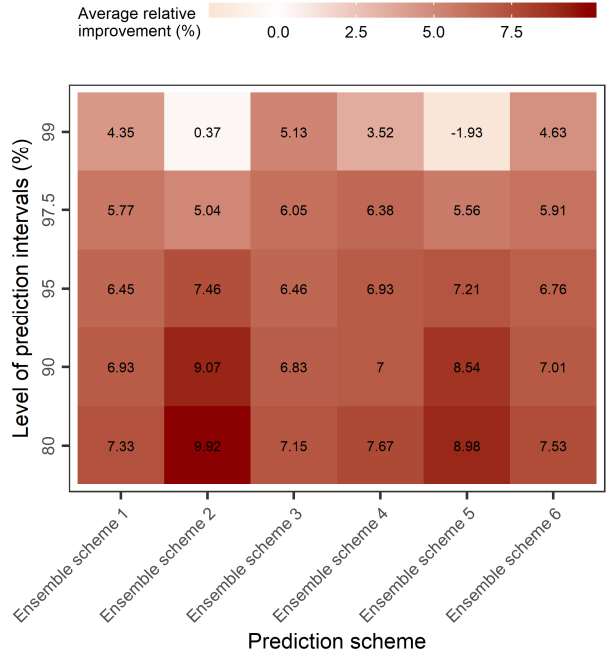

Figure D.2. Average relative improvements in terms of average interval score provided by the Bayesian calibration scheme with respect to the informal calibration scheme, when both these schemes are used as components of ensemble schemes 1–6. The latter are implemented with their remaining components and parameters set common. The legend limits are common for Figures D.2 and D.3. Each presented value summarizes 270 values.

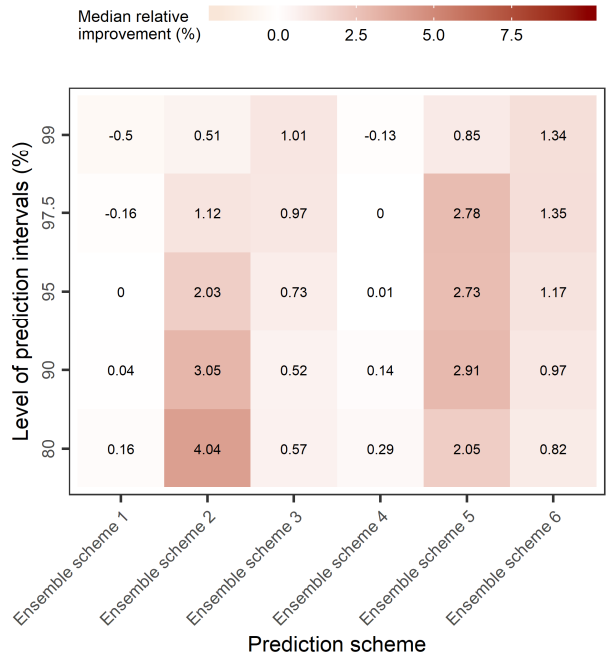

Figure D.3. Median relative improvements in terms of average interval score provided by the Bayesian calibration scheme with respect to the informal calibration scheme, when both these schemes are used as components of ensemble schemes 1–6. The latter are implemented with their remaining components and parameters set common. The legend limits are common for Figures D.2 and D.3. Each presented value summarizes 270 values.

**Appendix F   Additional remarks**

We have extensively explored through benchmark tests the modelling possibilities provided by the working methodology, when this methodology is applied for solving monthly rainfall-runoff problems using the quantile regression model as error model. Our benchmark experiment is of large-scale; nevertheless, it could not highlight all aspects of the working methodology. For exploiting this methodology in an optimal way, the following key adjustments to its components and parameters could be made:

o   The historical dataset can be divided in various ways, i.e., different proportions of the available information could be devoted to hydrological model calibration and error model training. This adjustment could be made to maximize predictive performance by exploiting evidence extracted from properly designed large-sample investigations. It could also be made for reducing the computational requirements, also depending on our choices on the remaining components and parameters. Applications to hundreds of catchments at timescales finer than the monthly one may require achieving a balance between predictive performance and computational requirements (when our computational resources are limited).

o   Any hydrological model (e.g., a process-based hydrological model of our preference) can be selected. Predictive performance improvements may be achieved by selecting one hydrological model over another or by adopting multi-model approaches (as proposed in Vrugt 2018, 2019, yet with the interest being in producing and combining quantile predictions instead of PDF predictions), thereby extending the working methodology, as suggested by Montanari and Koutsoyiannis (2012) for the original blueprint. Properly designed large-sample investigations could effectively guide our related choices.

o   The parameters of the hydrological model can be obtained by using a large variety of calibration schemes, including informal calibration schemes. (Note that random selection of the parameters, i.e., no period $T_1$, could also be an option). This point may be particular important for reducing the computational requirements. In Appendix E, we present large-sample investigations (on the monthly rainfall-runoff data exploited in the study) focusing on the comparison between Bayesian and informal calibration schemes for obtaining a large number of hydrological model parameters within the working methodology.

- The number of sister predictions can be selected based on the available computational resources. Nonetheless, the larger this number the larger the advantage of the methodology in terms of robustness (compared to basic two-stage post-processing methodologies). Properly designed benchmark experiments could also focus on optimizing this parameter of the working methodology (separately for the various timescales).
- Any statistical learning regression model that is suitable for predicting quantiles (e.g., the error models exploited in Papacharalampous et al. 2019d) can be selected as error model. This point may be particularly important for maximizing predictive performance (see also the key remarks in Section 4).
- Any set of predictor variables (e.g., the hydrological model predictions at times $t$, $t-1$, $t-2$, etc.) can be used in the application of the error model. This point may be important for maximizing predictive performance for timescales finer than the monthly one (see e.g., the findings in Papacharalampous et al. 2019d).
- All the above adjustments and modelling choices can be made separately for each of the three variants and for each level of prediction interval (or level of predictive quantile).

**Acknowledgements:** We sincerely thank the Editor, the Associate Editor, Dr Elena Volpi and an anonymous referee for their constructive reviews, which helped us to significantly improve the paper.

**Declarations of interest:** The authors declare no conflict of interest.

**Funding information:** The research work of Georgia Papacharalampous was supported by the Hellenic Foundation for Research and Innovation (HFRI) under the HFRI PhD Fellowship grant (Fellowship Number: 1388).